\documentclass[graybox]{svmult}

\usepackage{mathptmx}       
\usepackage{helvet}         
\usepackage{courier}        
\usepackage{type1cm}        
\usepackage{makeidx}         
\usepackage{graphicx}        
\usepackage{multicol}        
\usepackage[bottom]{footmisc}
\usepackage{rotating}
\usepackage{epsfig}
\usepackage{amssymb}

\newcommand{\be}{\begin{equation}}
\newcommand{\ee}{\end{equation}}
\newcommand{\bea}{\begin{eqnarray}}
\newcommand{\eea}{\end{eqnarray}}
\newcommand{\bnabla}{\mbox{\boldmath{$\nabla$}}}
\newcommand{\nn}{\nonumber}

\newcounter{saveeqn}%
\newcommand{\alpheqn}{\setcounter{saveeqn}{\value{equation}}%
\stepcounter{saveeqn}\setcounter{equation}{0}%
\renewcommand{\theequation}
  {\mbox{\arabic{saveeqn}\alph{equation}}}}%
\newcommand{\reseteqn}{\setcounter{equation}{\value{saveeqn}}%
\renewcommand{\theequation}{\arabic{equation}}}%

\makeindex             

\begin{document}

\title*{Local and Global Casimir Energies:  Divergences, Renormalization, and 
the Coupling to Gravity}
\titlerunning{Local and Global Casimir Energies} 
\author{Kimball A. Milton}
\institute{Kimball A. Milton \at Homer L. Dodge Department of Physics
and Astronomy, University of Oklahoma, Norman, OK 73019 \email{milton@nhn.ou.edu}}
%
%
\maketitle

\abstract{From the beginning of the subject, calculations of quantum vacuum 
energies or Casimir 
energies have been plagued with two types of divergences:  The total energy, 
which may be thought of as some sort of 
regularization of the zero-point energy, $\sum\frac12\hbar\omega$, seems 
manifestly divergent.  And local energy densities, obtained from the vacuum 
expectation 
value of the energy-momentum tensor, $\langle T_{00}\rangle$,  typically 
diverge near 
boundaries.  These two types of divergences have little to do with each other.
The energy of interaction between distinct rigid bodies of whatever type is 
finite, corresponding to observable forces and torques between the bodies, 
which can be unambiguously calculated.  The divergent local energy densities 
near surfaces do not change when the relative position of the rigid bodies is 
altered.  The self-energy 
of a body is less well-defined, and suffers divergences which may or may not 
be removable.
Some examples where a unique total self-stress may be evaluated include the 
perfectly conducting 
spherical shell first considered by Boyer, a perfectly conducting cylindrical 
shell, and 
dilute dielectric balls and cylinders.  In these cases the finite part 
is unique, yet 
there are divergent contributions which may be subsumed in some sort of 
renormalization 
of physical parameters.  The finiteness of self-energies is separate from the 
issue of 
the physical observability of the effect.  The divergences that occur 
in the local 
energy-momentum tensor near surfaces are distinct from the divergences in the 
total 
energy, which are often associated with energy located 
exactly on the surfaces.  
However, the local energy-momentum tensor couples to gravity, so what 
is the significance of infinite quantities here?  For the classic situation 
of parallel plates there are indications that the divergences in the local 
energy density are consistent with divergences in Einstein's equations;
correspondingly, it has been shown that divergences in the total Casimir
energy serve to precisely renormalize 
the masses of the plates, in accordance with the equivalence principle.  This 
should be a general property, but has not yet been established, for 
example, for the Boyer sphere. It is known that such local divergences 
can have no effect on macroscopic causality.}

\section{Introduction}
\label{sec:1}
\index{History of Casimir effect}
For more than 60 years it has been appreciated that quantum fluctuations
can give rise to macroscopic forces between bodies \cite{casimir}.  
These can be thought
of as the sum, in general nonlinear, of the van der Waals forces between the
constituents of the bodies, which, in the 1930s had been shown by London
\cite{london} 
to arise from dipole-dipole interactions in the nonretarded regime, and in
1947 to arise from the same interactions in the retarded regime, giving rise 
to so-called Casimir-Polder forces
\cite{casimirandpolder}.  Bohr \cite{casimir50} 
apparently provided the incentive
to Casimir to rederive the macroscopic force between a molecule and a 
surface, and then derive the force between two conducting surfaces, directly
in terms of zero-point fluctuations of the electromagnetic fields in which 
the bodies are immersed.  But these two points of view---action at a distance
and local action---are essentially equivalent, and one implies the other,
not withstanding some objections to the latter \cite{Jaffe:2003ji}.

The quantum-vacuum-fluctuation force between two parallel surfaces---be they
conductors or dielectrics \cite{lifshitz,dzyaloshinskii0,dzyaloshinskii}
---was the first situation considered, and still
the only one accessible experimentally. 
(For a current review of the experimental situation, see
\cite{Bordag:2009zz,Klimchitskaya:2009cw})
 Actually, most experiments measure
the force between a spherical surface and a plane, but the surfaces are so
close together that the force may be obtained from the parallel plate case
by a geometrical transformation, the so-called proximity force approximation
(PFA) \cite{derpt,derpt2,blocki}.  
However, it is not possible to find an extension to the PFA beyond
the first approximation of the separation distance being smaller than all
other scales in the problem.
In the last few years, advances in technique have allowed quasi-analytical
and numerical
calculations to be carried out between bodies of essentially any shape, at
least at medium to large separation, so the limitations of the PFA may be
largely transcended.  (For the current status of these developments,
see the contributions to this volume by Emig, Jaffe, and Rahi, and by
Johnson; for earlier references, see, for example \cite{Milton:2008st}.)
These advances have shifted calculational attention
away from what used to be the central challenge in Casimir theory, how to
define and calculate Casimir energies and self-stresses of single bodies.

There are, of course, sound reasons for this.  Forces between distinct
bodies are necessarily physically finite, and can, and have, been observed
by experiment.  Self-energies or self-stresses typically involve divergent
quantities which are difficult to remove, and have obscure physical meaning.
For example, the self-stress on a perfectly conducting spherical shell of
negligible thickness was calculated by Boyer in 1968 \cite{Boyer:1968uf}, 
who found a repulsive
self-stress  that has subsequently been confirmed by a variety of techniques.
Yet it remains unclear what physical significance this energy has.  If
the sphere is bisected and the two halves pulled apart, there will be
an attraction (due to the closest parts of the hemispheres) not a repulsion.
The same remarks, although exacerbated, apply to the self-stress on a
rectangular box \cite{lukosz,lukosz1,lukosz2,ambjorn}.
  The situation in that case is worse because (1) the
sharp corners give rise to additional divergences not present in the
case of a smooth boundary (it has been proven that the self-energy of
a smooth closed infinitesimally thin conducting surface is finite 
\cite{balian,Bernasconi}), and (2) the
exterior contributions cannot be computed because the vector Helmholtz
equation cannot be separated.  But calculational challenges aside,
the physical significance of self-energy remains elusive.

The exception to this objection is provided by gravity.  Gravity
couples to the local energy-momentum or stress tensor, and, in the
leading quantum approximation, it is the vacuum expectation value
of the stress tensor that provides the source term in Einstein's equations.
Self energies should therefore in principle be observable.  This
is largely uncharted territory, except in the instance of the classic
situation of parallel plates.  There, after a bit of initial confusion,
it has now been established that the divergent self-energies of each
plate in a two-plate apparatus, as well as the mutual Casimir energy
due to both plates, gravitates according to the equivalence principle,
so that indeed it is consistent to absorb the divergent self-energies of
each plate into the gravitational and inertial mass of each
\cite{Fulling:2007xa,Milton:2007ar}.
This should be a universal feature.

In this paper, for pedagogical reasons, we will concentrate attention
on the Casimir effect due to massless scalar field fluctuations, where
the potentials are described by $\delta$-function potentials, so-called
semitransparent boundaries.  In the limit as the coupling to these potentials
becomes infinitely strong, this imposes Dirichlet boundary conditions.
At least in some cases, Neumann boundary conditions can be achieved
by the strong coupling limit of the derivative of $\delta$-function
potentials.  So we can, for planes, spheres, and circular cylinders,
recover in this way the results for electromagnetic field fluctuations
imposed by perfectly conducting boundaries.  Since the mutual interaction
between distinct semitransparent bodies have been described in detail
elsewhere \cite{Milton:2007gy,Milton:2007wz,Wagner:2008qq}, we will,
as implied above, concentrate on the self-interaction issues.

A summary of what is known for spheres and circular cylinders is
given in Table \ref{tab1}.\index{Casimir energies for spheres and cylinders}

\begin{table}
\caption{Casimir energy ($E$) for a sphere and Casimir energy per unit
length ($\mathcal{E}$) for a cylinder, both of radius $a$.
Here the different boundary
conditions are perfectly conducting for electromagnetic fields (EM),
Dirichlet for scalar fields (D), dilute dielectric for electromagnetic
fields [coefficient of $(\varepsilon-1)^2$], dilute dielectric for
electromagnetic fields
with media having the same speed of light (coefficient of $\xi^2
=[(\varepsilon-1)/(\varepsilon+1)]^2$),
 perfectly conducting surface with eccentricity $\delta e$
(coefficient of $\delta e^2$), and weak coupling
for scalar field with $\delta$-function boundary given by (\ref{deltapot}),
(coefficient of $\lambda^2/a^2$).  The references given are, to the author's
knowledge, the
first paper in which the results in the various cases were found.}
\label{tab1}
\begin{tabular}{p{2cm}p{2.4cm}p{2cm}p{4.9cm}}
\hline\noalign{\smallskip}
Type&$E_{\rm Sphere}a$&$\mathcal{E}_{\rm Cylinder}a^2$&References\\
\noalign{\smallskip}\svhline\noalign{\smallskip}
EM&$+0.04618$&$-0.01356$&\cite{Boyer:1968uf} \cite{DeRaad:1981hb}\\
D&$+0.002817$&$+0.0006148$&\cite{Bender:1994zr}\cite{gosrom}\\
$(\varepsilon-1)^2$&$+0.004767=\frac{23}{1536\pi}$&$0$
&\cite{Brevik:1998zs}\cite{Cavero-Pelaez:2004xp}\\
$\xi^2$&$+0.04974=\frac5{32\pi}$&$0$&\cite{Klich:1999df}\cite{nesterenko}\\
$\delta e^2$&$\pm0.0009$&$0$&\cite{Kitson:2005kk}\cite{kitsonromeo}\\
$\lambda^2/a^2$&$+0.009947=\frac1{32\pi}$&$0$&\cite{Milton:2002vm}\cite{CaveroPelaez:2006rt}\\
\noalign{\smallskip}\hline\noalign{\smallskip}
\end{tabular}

\end{table}

\section{Casimir Effect Between Parallel Plates: A $\delta$-Potential
Derivation}\label{Sec1}
\index{Casimir effect between parallel plates}
In this section, we will rederive the classic Casimir result for the
force between parallel conducting plates \cite{casimir}.  Since the
usual Green's function derivation may be found in monographs \cite{miltonbook},
and was for example reviewed in connection with current controversies over
finiteness of Casimir energies \cite{Milton:2002vm}, we will here present
a different approach, based on $\delta$-function potentials, which in the
limit of strong coupling reduce to the appropriate Dirichlet or Robin
boundary conditions of a perfectly conducting surface, as appropriate to
TE and TM modes, respectively.  Such potentials were first considered
by the Leipzig group  \cite{hennig,bkv}, but more recently have been the focus
of the program of the MIT group \cite{graham,graham2,Graham:2002fw,Graham:2003ib}.
The discussion here is based on a paper by the author
\cite{Milton:2004vy}. (See also \cite{Milton:2004ya}.)  
(A multiple scattering approach to this problem has also
been given in \cite{Milton:2007wz}.)

We consider a massive scalar field (mass $\mu$)
 interacting with two $\delta$-function
potentials, one at $x=0$ and one at $x=a$, which has an interaction
Lagrange density
\be
\mathcal{L}_{\rm int}=-\frac12\lambda\delta(x)\phi^2(x)
-\frac12\lambda'\delta(x-a)\phi^2(x),
\label{oneplusonelag}
\ee
\index{$\delta$-function potentials}
where  the positive coupling constants $\lambda$ and $\lambda'$
have dimensions of mass.  In the limit as both
couplings become infinite, these potentials enforce Dirichlet boundary
conditions at the two points:
\be
\lambda,\lambda'\to\infty:\qquad \phi(0),\phi(a)\to0.
\ee

The Casimir energy for this
situation may be computed in terms of the Green's function $G$,
\be
G(x,x')=\I\langle T\phi(x)\phi(x')\rangle,
\label{fgf}
\ee
which has a time Fourier transform,
\be
G(x,x')=\int\frac{\D\omega}{2\pi}\E^{-\I\omega(t-t')}\mathcal{G}(x,x';\omega).
\label{ft}
\ee
Actually, this is a somewhat symbolic expression, for the Feynman Green's
function (\ref{fgf}) implies that the frequency contour of integration
here must pass below the singularities in $\omega$ on the negative real
axis, and above those on the positive real axis \cite{kantowski,Brevik:2000hk}.
Because we have translational invariance in the two directions parallel
to the plates, we have a Fourier transform in those directions as well:
\be
\mathcal{G}(x,x';\omega)=\int\frac{(\D\mathbf{k})}
{(2\pi)^2}\E^{\I\mathbf{k\cdot(r-r')_\perp}}g(x,x';\kappa),
\label{redgreen}
\ee
where $\kappa^2=\mu^2+k^2-\omega^2$.  

The reduced Green's function in (\ref{redgreen}) in turn satisfies
\be
\left[-\frac{\partial^2}{\partial x^2}+\kappa^2+\lambda\delta(x)
+\lambda'\delta(x-a)\right]g(x,x')=\delta(x-x').
\ee
This equation is easily solved, with the result
\alpheqn
\label{gee0}
\bea
g(x,x')&=&\frac1{2\kappa}e^{-\kappa|x-x'|}+
\frac1{2\kappa\Delta}\Bigg[\
\frac{\lambda\lambda'}{(2\kappa)^2}2\cosh\kappa|x-x'|\nonumber\\
&&\quad\mbox{}-\frac{\lambda}{2\kappa }\left(1+\frac{\lambda'}
{2\kappa }\right)\E^{2\kappa a}
\E^{-\kappa(x+x')}-\frac{\lambda'}{2\kappa}\left(1+\frac{\lambda}
{2\kappa}\right)\E^{\kappa(x+x')}\Bigg]
\label{gin}
\eea
for both fields inside, $0<x,x'<a$, while if both field points are outside,
$a<x,x'$,
\bea
 g(x,x')&=&\frac1{2\kappa}\E^{-\kappa|x-x'|}+\frac1{2\kappa\Delta}\E^{-\kappa
(x+x'-2a)}\nn\\
&&\quad\times\left[-\frac{\lambda}{2\kappa}\left(1-\frac{\lambda'}
{2\kappa}\right)
-\frac{\lambda'}{2\kappa}\left(1+\frac{\lambda}
{2\kappa}\right)\E^{2\kappa a}\right].
\label{gout}
\eea
For $x,x'<0$,
\bea
 g(x,x')&=&\frac1{2\kappa}\E^{-\kappa|x-x'|}+\frac1{2\kappa\Delta}
\E^{\kappa(x+x')}\nn\\
&&\quad\times\left[-\frac{\lambda'}{2\kappa}\left(1-\frac{\lambda}
{2\kappa}\right)
-\frac{\lambda}{2\kappa}\left(1+\frac{\lambda'}
{2\kappa}\right)\E^{2\kappa a}\right].
\label{gleft}
\eea
\reseteqn
Here, the denominator is
\be
\Delta=\left(1+\frac{\lambda}{2\kappa}\right)\left(1+\frac{\lambda'}
{2\kappa}\right)\E^{2\kappa a}-\frac{\lambda\lambda'}{(2\kappa)^2}.
\label{delta0}
\ee
Note that in the strong coupling limit we recover the familiar results,
for example, inside
\be
\lambda,\lambda'\to\infty:
\quad g(x,x')\to-\frac{\sinh\kappa x_<\sinh\kappa(x_>-a)}
{\kappa\sinh\kappa a}.
\ee
Here $x_>$, $x_<$ denote the greater, lesser, of $x,x'$.
Evidently, this Green's function vanishes at $x=0$ and at $x=a$.

Let us henceforward consider $\mu=0$, since otherwise there are
no long-range forces.  (There is no nonrelativistic Casimir effect.)
We can now calculate the force on one of the $\delta$-function plates by
calculating the discontinuity of the stress tensor,  obtained from the
Green's function (\ref{fgf}) by
\be
\langle T^{\mu\nu}\rangle=\left(\partial^\mu\partial^{\nu\prime}-\frac12
g^{\mu\nu}\partial^\lambda\partial'_\lambda\right)
\frac1{\I}G(x,x')\bigg|_{x=x'}.
\ee
Writing a reduced stress tensor by
\be
\langle T^{\mu\nu}\rangle=\int\frac{\D\omega}{2\pi}
\int\frac{(\D\mathbf{k})}{(2\pi)^2} t^{\mu\nu},
\ee
we find inside, just to the left of the plate at $x=a$,
\alpheqn
\index{Stress tensor}
\bea
t_{xx}\big|_{x=a-}&=&\frac1{2\I}(-\kappa^2+\partial_x\partial_{x'})g(x,x')
\bigg|_{x=x'=a-}\label{txxin}\\
&=&-\frac\kappa{2\I}\left\{
1+2\frac{\lambda\lambda'}{(2\kappa)^2}\frac1\Delta\right\}.
\label{stressin}
\eea
\reseteqn
From this we must subtract the stress just to the right of the plate at
$x=a$, obtained from (\ref{gout}), which turns out to be in the massless
limit
\be
t_{xx}\big|_{x=a+}=-\frac\kappa{2\I},\label{stressout}
\ee
which just cancels the 1 in braces in (\ref{stressin}).
Thus the pressure on the plate at $x=a$ due to the quantum fluctuations 
in the scalar field is given by the simple, finite expression
\be
 P=\langle T_{xx}\rangle\big|_{x=a-}-\langle T_{xx}\rangle\big|_{x=a+}
=-\frac1{32\pi^2 a^4}\int_0^\infty
\D y\,y^2\,\frac1{(y/(\lambda a)+1)(y/(\lambda'a)+1)\E^y-1},
\label{ceplates}
\ee
which coincides with the result given in
\cite{Graham:2003ib,Weigel:2003tp}.  The leading behavior
for small $\lambda=\lambda'$ is
\alpheqn
\be
P^{\rm TE}\sim-\frac{\lambda^2}{32\pi^2 a^2},\qquad \lambda\ll 1,
\label{pertte}
\ee
while for large $\lambda$ it approaches half of Casimir's result \cite{casimir}
for perfectly conducting parallel plates,
\be
P^{\rm TE}\sim -\frac{\pi^2}{480 a^4},\qquad\lambda\gg1.
\label{strongte}
\ee
\reseteqn


We can also compute the energy density.  Integrating the energy density
over all space should give rise to the total energy.
Indeed, the above result may be easily
derived from the following expression for the total energy,
\bea
E=\int(\D\mathbf{r})\,\langle T^{00}\rangle&=&\frac1{2\I}\int(\D\mathbf{r})
(\partial^0\partial^{\prime0}-\nabla^2)G(x,x')\bigg|_{x=x'}\nonumber\\
&=&\frac1{2\I}\int(\D\mathbf{r})\int\frac{\D\omega}{2\pi}
2\omega^2\mathcal{G}(\mathbf{r,r}),
\label{casenergy}
\eea
if we integrate by parts and omit the surface term.
Integrating over the Green's functions in the three regions,
given by (\ref{gin}), (\ref{gout}), and (\ref{gleft}), we obtain for
$\lambda=\lambda'$,
\be
\mathcal{E}=\frac1{48\pi^2 a^3}\int_0^\infty \D y\,y^2
\frac1{1+y/(\lambda a)}-\frac1{96\pi^2 a^3}
\int_0^\infty \D y\,y^3\frac{1+2/(y+\lambda a)}{(y/(\lambda a)+1)^2\E^y-1},
\label{31energy}
\ee
where the first term is regarded as an irrelevant constant ($\lambda$ is
constant so the $a$ can be scaled out), 
and the second term
coincides with the massless limit of the energy first found by Bordag
et al.~\cite{hennig}, and given in \cite{Graham:2003ib,Weigel:2003tp}.
When differentiated with respect to $a$,
(\ref{31energy}), with $\lambda$ fixed, yields the pressure
(\ref{ceplates}).  (We will see below that the divergent constant
describe the self-energies of the two plates.)

If, however, we integrate the interior and exterior energy density
directly, one gets a different result.
The origin of this discrepancy with the naive energy
is the existence of a surface contribution
to the energy. To see this, we must include the potential in the stress
tensor,
\be
T^{\mu\nu}=\partial^\mu\phi\partial^\nu\phi-\frac12 g^{\mu\nu}
\left(\partial^\lambda\phi\partial_\lambda\phi+V\phi^2\right),
\ee
and then, using the equation of motion, it is immediate to see
that the energy density is
\be
T^{00}=\frac12\partial^0\phi\partial^0\phi-\frac12\phi(\partial^0)^2\phi
+\frac12\bnabla\cdot(\phi\bnabla\phi),\label{t00pluseom}
\ee
so, because the first two terms here yield the last form in
(\ref{casenergy}), we conclude that there is
an additional contribution to the energy,
\alpheqn
\bea
\hat E&=&-\frac1{2\I}\int \D\mathbf{S}\cdot\bnabla G(x,x')\bigg|_{x'=x}
\label{es1}\\
&=&-\frac1{2\I}\int_{-\infty}^\infty\frac{\D\omega}{2\pi}
\int\frac{(\D\mathbf{k})}{(2\pi)^2}\sum\frac{\D}{\D x}
g(x,x')\bigg|_{x'=x},
\label{es2}
\eea
\reseteqn
\index{Surface energy}
where the derivative is taken at the boundaries (here $x=0$, $a$) in the
sense of the outward normal from the region in question.  When this surface
term is taken into account the extra terms incorporated in (\ref{31energy})
are supplied.  The integrated formula (\ref{casenergy})
automatically builds in this
surface contribution, as the implicit surface term in the integration
by parts.  That is,
\be
E=\int(\D{\bf r})\langle T^{00}\rangle+\hat E.\label{einclst}
\ee
(These terms are slightly unfamiliar because they do not arise
in cases of Neumann or Dirichlet boundary conditions.)  See Fulling
\cite{Fulling:2003zx} for further discussion.  That the surface
energy of an interface arises from the volume energy of a smoothed
interface is demonstrated in \cite{Milton:2004vy}, and elaborated
in Sect.~\ref{sec:2.4}.


In the limit of strong coupling, we obtain
\be
\lim_{\lambda\to\infty}\mathcal{E}=-\frac{\pi^2}{1440 a^3},
\label{1/2casimir}
\ee
which is exactly one-half the energy found by Casimir for
perfectly conducting plates \cite{casimir}.
Evidently, in this case, the TE modes (calculated here) and
the TM modes (calculated in the following subsection) give equal contributions.

\subsection{TM Modes}
\label{sec:tm11}
\index{Transverse magnetic modes}
To verify this last claim, we solve a similar problem with boundary conditions
that the derivative of $g$ is continuous at $x=0$ and $a$,
\alpheqn
\be
\frac\partial{\partial x}g(x,x')\bigg|_{x=0,a} \mbox{ is continuous},
\label{tmbc1}
\ee
but the function itself is discontinuous,
\be
g(x,x')\bigg|_{x=a-}^{x=a+}=\lambda  \frac\partial{\partial x}g(x, x')\bigg|
_{x=a},
\label{tmbc2}
\ee
and similarly at $x=0$.  (Here the coupling $\lambda$ has dimensions of length.)
These boundary conditions reduce, in the limit of strong coupling, to
Neumann boundary conditions on the planes, appropriate to electromagnetic
TM modes:
\be
\lambda\to\infty:\qquad\frac\partial{\partial x}g(x,x')\bigg|_{x=0,a}=0.
\label{nbc}
\ee
\reseteqn
It is completely straightforward to work out the reduced Green's function
in this case.  When both points are between the planes, $0<x,x'<a$,
\alpheqn
\bea
g(x,x')=\frac1{2\kappa}\E^{-\kappa|x-x'|}+\frac1{2\kappa\tilde \Delta}
\Bigg\{\left(\frac{\lambda  \kappa }2\right)^2 2\cosh\kappa(x-x')\nonumber\\
\mbox{}+
\frac{\lambda \kappa }2\left(1+\frac{\lambda \kappa }2\right)
\left[
\E^{\kappa(x+x')}+\E^{-\kappa(x+x'-2a)}\right]\Bigg\},
\eea
while if both points are outside the planes, $a<x,x'$,
\bea
g(x,x')&=&\frac1{2\kappa}\E^{-\kappa|x-x'|}\nonumber\\
&&\mbox{}+\frac1{2\kappa\tilde\Delta}\frac{\lambda \kappa }2
\E^{-\kappa(x+x'-2a)}\left[\left(1-\frac{\lambda \kappa }2\right)+
\left(1+\frac{\lambda \kappa }2\right)\E^{2\kappa a}\right],
\eea
\reseteqn
where the denominator is
\be
\tilde\Delta=\left(1+\frac{\lambda \kappa }2\right)^2\E^{2\kappa a}-
\left(\frac{\lambda \kappa }2\right)^2.
\ee

It is easy to check that in the strong-coupling limit, the appropriate
Neumann boundary condition (\ref{nbc}) is recovered.  For example, in the
interior region, $0<x,x'<a$,
\be
\lim_{\lambda\to\infty}g(x,x')=\frac{\cosh\kappa x_<\cosh\kappa(x_>-a)}{\kappa
\sinh\kappa a}.
\ee

\begin{figure}
\sidecaption
\begin{turn}{270}
\epsfig{figure=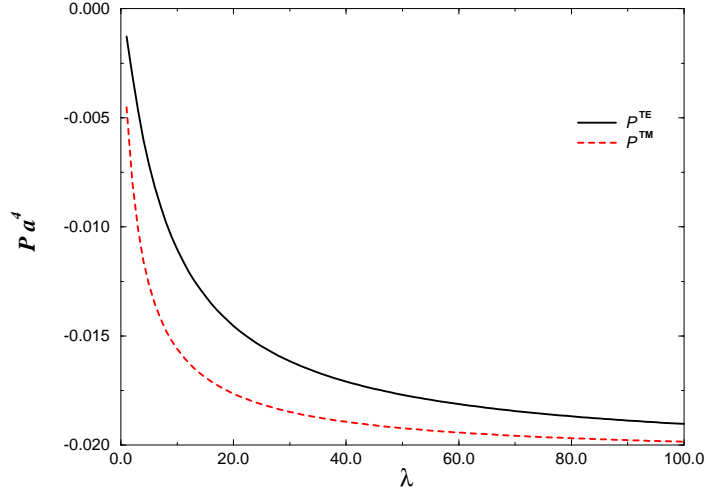,height=10cm}
\end{turn}
\caption{TE and TM Casimir pressures between
 $\delta$-function planes having strength
$\lambda$ and separated by a distance $a$.
In each case, the pressure is plotted as a function
of the dimensionless coupling, $\lambda a$ or $\lambda/a$,
respectively, for TE and TM contributions.}
\label{fig2}
\end{figure}

Now we can compute the pressure on the plane by computing the $xx$ component
of the stress tensor,  which is given by (\ref{txxin}),
so we find
\alpheqn
\bea
t_{xx}\big|_{x=a-}&=&\frac1{2\I}\left[-\kappa-\frac{2\kappa}{\tilde\Delta}
\left(\frac{\lambda \kappa }2\right)^2\right],\\
t_{xx}\big|_{x=a+}&=&-\frac1{2\I}\kappa,
\eea
\reseteqn
and the flux of momentum deposited in the plane $x=a$ is
\be
t_{xx}\big|_{x=a-}-t_{xx}\big|_{x=a+}=\frac{\I \kappa}{\left(\frac2{\lambda
\kappa }+1\right)^2\E^{2\kappa a}-1},
\ee
and then by integrating over frequency and transverse momentum we obtain
the pressure:
\be
P^{\rm TM}=-\frac1{32\pi^2 a^4}\int_0^\infty \D y\,y^3\frac1{
\left(\frac{4a}{\lambda  y}+1\right)^2\E^y-1}.
\label{ptm}
\ee
In the limit of weak coupling, this behaves as follows:
\be
P^{\rm TM}\sim -\frac{15}{64\pi^2 a^6}\lambda^2,
\ee
which is to be compared with (\ref{pertte}).
In strong coupling, on the other hand, it has precisely the same limit as
the TE contribution, (\ref{strongte}), which confirms the expectation
given at the end of the previous subsection.  Graphs of the two functions
are given in Fig.~\ref{fig2}.

For calibration purposes we give the Casimir pressure in practical units
between ideal perfectly conducting parallel plates at zero temperature:
\index{Casimir pressure for perfectly conducting parallel plates}
\be
P=-\frac{\pi^2}{240 a^4}\hbar c=-\frac{1.30\mbox{ mPa}}{(a/1\mu\mbox{m})^4}.
\label{casresult}
\ee

\subsection{Self-energy of Boundary Layer}
\label{sec:2.4}
\index{Self energy}
Here we show that the divergent self-energy of a single plate,
half the divergent term in (\ref{31energy}),
 can be interpreted as the energy associated 
with the boundary layer.  We do this in a simple
context by considering a scalar field
interacting with the background
\be
\mathcal{L}_{\rm int}=-\frac\lambda 2\phi^2\sigma,
\label{bkgdpot}
\ee
where the background field $\sigma$  expands the meaning of the
$\delta$ function,
\be
\sigma(x)=\left\{\begin{array}{cc}
h,&-\frac\delta2<x<\frac\delta 2,\\
0,&\mbox{otherwise},
\end{array}\right.
\ee
with the property that $h\delta=1$.
The reduced Green's function satisfies
\be
\left[-\frac{\partial^2}{\partial x^2}
+\kappa^2+\lambda\sigma(x)\right]g(x,x')=\delta(x-x').
\ee
This may be easily solved in the region of the slab, $-\frac\delta2<x<\frac
\delta2$,
\bea
g(x,x')=\frac1{2\kappa'}\bigg\{\E^{-\kappa'|x-x'|}
+\frac1{\hat\Delta}\bigg[\lambda h\cosh\kappa'(x+x')\nonumber\\
\qquad\mbox{}+(\kappa'-\kappa)^2\E^{-\kappa'\delta}\cosh\kappa'(x-x')\bigg]
\bigg\}.
\label{slabgf}
\eea
Here $\kappa'=\sqrt{\kappa^2+\lambda h}$, and
\be
\hat\Delta=2\kappa\kappa'\cosh\kappa'\delta+
(\kappa^2+\kappa^{\prime2})\sinh\kappa'\delta.
\ee
This result may also easily be derived from the multiple reflection
formulas given in \cite{Milton:2004ya}, and agrees with that given
by Graham and Olum \cite{Graham:2002yr}. 

Let us
proceed here with more generality, and consider the stress tensor with
an arbitrary conformal term \cite{ccj},
\be
T^{\mu\nu}=\partial^\mu\phi\partial^\nu\phi-\frac12 g^{\mu\nu}(\partial_\lambda
\phi\partial^\lambda\phi+\lambda h\phi^2)-\xi(\partial^\mu\partial^\nu
-g^{\mu\nu}\partial^2)\phi^2,\label{stconformal}
\ee
in $d+2$ dimensions, $d$ being the number of transverse dimensions,
and $\xi$ is an arbitrary parameter, sometimes called the conformal
parameter.\index{Conformal stress tensor}
Applying the corresponding differential operator to the Green's function
(\ref{slabgf}),  introducing polar coordinates in the $(\zeta,k)$ plane,
with $\zeta=\kappa\cos\theta$, $k=\kappa\sin\theta$, and
\be
\langle\sin^2\theta\rangle=\frac{d}{d+1},
\ee
we get the following form for the energy density within the slab.
\bea
\langle T^{00}\rangle &=&\frac{2^{-d-2}\pi^{-(d+1)/2}}
{\Gamma((d+3)/2)}\int_0^\infty \frac{\D
\kappa\,\kappa^d}{\kappa'\hat\Delta}\bigg\{\lambda h
\left[(1-4\xi)(1+d)\kappa^{\prime2}-\kappa^2\right]\cosh2\kappa'x\nonumber\\
&&\mbox{}-(\kappa'-\kappa)^2\E^{-\kappa'\delta}\kappa^2\bigg\},
\quad -\delta/2<x<\delta/2.\label{t00slab}
\eea
We can also calculate the energy density on the other side of the boundary,
from the Green's function for $x,x'<-\delta/2$,
\be
g(x,x')=\frac1{2\kappa}\left[\E^{-\kappa|x-x'|}-\E^{\kappa(x+x'+\delta)}
\lambda h\frac{\sinh\kappa'\delta}{\hat\Delta}\right],
\ee
and the corresponding energy density is given by
\be
\langle T^{00} \rangle=-\frac{d(1-4\xi (d+1)/d)}{2^{d+2}\pi^{(d+1)/2}\Gamma((d+3)/2)}
\int_0^\infty \D \kappa\,\kappa^{d+1}\frac1{\hat\Delta}\lambda h
\E^{2\kappa (x+\delta/2)}\sinh\kappa'\delta,\label{t00out}
\ee
which vanishes if the conformal value of $\xi$ is used.  An identical
contribution comes from the region $x>\delta/2$.

Integrating $\langle T^{00}\rangle$ 
over all space gives the vacuum energy of the slab
\bea
E_{\rm slab}&=&-\frac1{2^{d+2}\pi^{(d+1)/2}\Gamma((d+3)/2)}
\int_{0}^\infty \D\kappa\,\kappa^d\frac1{\kappa'\hat
\Delta}\bigg[(\kappa'-\kappa)^2\kappa^2
\E^{-\kappa'\delta}\delta\nonumber\\
&&\quad\mbox{}+(\lambda h)^2
\frac{\sinh\kappa'\delta}{\kappa'}\bigg].
\eea
Note that the conformal term does not contribute to the total energy.
If we now take the limit $\delta\to0$ and $h\to \infty$ so that $h\delta=1$,
we immediately obtain the self-energy of a single $\delta$-function plate:
%
\be E_\delta=
\lim_{h\to\infty}E_{\rm slab}=\frac1{2^{d+2}\pi^{(d+1)/2}\Gamma((d+3)/2)}
\int_0^\infty \D \kappa \,\kappa^d\frac\lambda{\lambda+2\kappa}.\label{eslab}
\ee
which for $d=2$ precisely coincides with one-half the constant term in 
(\ref{31energy}).

There is no surface term in the total Casimir energy as long as the
slab is of finite width, because we may easily check that $\frac{d}{dx}g
\big|_{x=x'}$ is continuous at the boundaries $\pm\frac\delta2$.  However,
if we only consider the energy internal to the slab we encounter not
only the integrated energy density but a surface term from the integration
by parts---see (\ref{einclst}). 
 It is the complement of this boundary term that gives rise to $E_{\delta}$,
(\ref{eslab}),
in this way of proceeding. That is, as $\delta\to0$, 
\be
-\int_{\rm slab}(d\mathbf{r})\int d\zeta\,\zeta^2\mathcal{G}(\mathbf{r,r})=0,
\ee
so
\be
E_\delta=\hat E\big|_{x=-\delta/2}+\hat E\big|_{x=\delta/2},
\ee with the normal defining the surface energies pointing into the slab.
This means that in this limit, the slab and surface energies coincide.
\index{Surface energy}

\index{Divergences in local energy}
Further insight is provided by examining the local energy density.
In this we follow the work of Graham and Olum \cite{Graham:2002yr,Olum:2002ra}.
From (\ref{t00slab}) we can calculate the behavior of the energy density as the
boundary is approached from the inside:
\be
\langle T^{00}\rangle \sim \frac{\Gamma(d+1)\lambda h}{2^{d+4}\pi^{(d+1)/2}\Gamma((d+3)/2)}
\frac{1-4\xi(d+1)/d}{(\delta-2|x|)^d},\quad |x|\to\delta/2.
\label{slabsurfdiv}
\ee  For $d=2$ for example, this agrees with the result found in
\cite{Graham:2002yr} for $\xi=0$:
\be
\langle T^{00}\rangle\sim\frac{\lambda h}{96\pi^2}\frac{(1-6\xi)}{(\delta/2-|x|)^2},
\qquad|x|\to\frac\delta2.\label{surfdiv}
\ee
Note that, as we expect, this surface divergence vanishes for the conformal
stress tensor \cite{ccj}, where $\xi=d/4(d+1)$.  (There will be subleading
divergences if $d>2$.)
The divergent term in the local energy density from the outside,
(\ref{t00out}), as $x\to-\delta/2$, is just the negative of that found in
(\ref{slabsurfdiv}).  This is why, when the total energy is computed by
integrating the energy density, it is  finite for $d<2$, and independent
of $\xi$. The divergence encountered for $d=2$ may be handled by
renormalization of the interaction potential \cite{Graham:2002yr}.

Note, further, that for a thin slab, close to the exterior but
such that the slab still appears thin,
$x\gg \delta$,  the sum of the exterior and interior energy density divergences
combine to give the energy density outside a $\delta$-function potential:
\be 
u_{\delta}=-\frac\lambda{96\pi^2}(1-6\xi)\left[\frac{h}{(x-\delta/2)^2}-
\frac{h}{(x+\delta/2)^2}\right]=
-\frac\lambda{48\pi^2}\frac{1-6\xi}{x^3},\label{pertsd}
\ee for small $x$. 
Although this limit might be criticized as illegitimate, this result
is correct for a $\delta$-function potential, and
we will see that this divergence structure occurs also in
spherical and cylindrical geometries, so that it is a universal surface
divergence without physical significance, barring gravity.
 
For further discussion on surface divergences, see Sect.~\ref{sec:surfdiv}.

\section{Surface and Volume Divergences}
\label{sec:surfdiv}
\index{Surface and volume divergences}

It is well known as we have just seen that in general the Casimir energy
density diverges in the neighborhood of a surface.  For flat surfaces
and conformal theories (such as the conformal scalar theory considered
above \cite{Milton:2002vm}, or electromagnetism) those divergences are not
present.\footnote{In general, this need not be the case.  For example,
Romeo and Saharian \cite{Romeo:2000wt} show that with mixed boundary conditions
the surface divergences need not vanish for parallel plates.
For additional work on local effects with mixed (Robin)
boundary conditions, applied to spheres and cylinders, and corresponding
global effects, see \cite{Romeo:2001dd,Saharian:2000mw,Saharian:2003dr,Fulling:2003zx}.
See also Sect.~\ref{sec:2.4} and \cite{Graham:2002yr,Olum:2002ra}.}
  In particular, Brown and Maclay \cite{Brown:1969na}
calculated the local stress tensor for two ideal plates separated by a distance
$a$ along the $z$ axis, with the result for a conformal scalar
\be
\langle T^{\mu\nu}\rangle
=-\frac{\pi^2}{1440 a^4}[4\hat z^\mu\hat z^\nu-g^{\mu\nu}].
\ee
This result was given more recent rederivations in
\cite{Actor:1996zj,Milton:2002vm}.
Dowker and Kennedy \cite{dowkerandkennedy} and Deutsch and
Candelas \cite{deutsch} considered the local
stress tensor between planes inclined at an angle $\alpha$,
with the result, in cylindrical coordinates $(t,r,\theta,z)$,
\be
\langle T^{\mu\nu}\rangle=-\frac{f(\alpha)}{720 \pi^2r^4}\left(
\begin{array}{cccc}
1&0&0&0\\
0&-1&0&0\\
0&0&3&0\\
0&0&0&-1
\end{array}\right),
\ee
where for a conformal scalar, with Dirichlet boundary conditions,
\be
f(\alpha)=\frac{\pi^2}{2\alpha^2}\left(\frac{\pi^2}{\alpha^2}
-\frac{\alpha^2}{\pi^2}\right),
\ee
and for electromagnetism, with perfect conductor boundary conditions,
\be
f(\alpha)=\left(\frac{\pi^2}{\alpha^2}+11\right)\left(\frac{\pi^2}{\alpha^2}-1
\right).
\ee
For $\alpha\to0$ we recover the pressures and energies for parallel plates,
(\ref{strongte}) and (\ref{casresult}).
(These results were later discussed in \cite{brevly}.)

Although for perfectly conducting
flat surfaces, the energy density is finite, for electromagnetism
the individual electric and magnetic fields have divergent RMS values,
\be
\langle E^2\rangle\sim-\langle B^2\rangle\sim\frac1{\epsilon^4},
\qquad\epsilon\to0,
\ee
a distance $\epsilon$ above a conducting surface.  However, if the surface
is a dielectric, characterized by a plasma dispersion relation,
 these divergences are softened
\be
\langle E^2\rangle\sim\frac1{\epsilon^3},\qquad
-\langle B^2\rangle\sim\frac1{\epsilon^2},\qquad\epsilon\to0,
\ee
so that the energy density also diverges \cite{sopovaqfext,Sopova:2005sx}
\be
\langle T^{00}\rangle\sim\frac1{\epsilon^3},\qquad\epsilon\to0.
\ee
The null energy condition ($n_\mu n^\mu=0$)
\be
T^{\mu\nu}n_\mu n_\nu\ge0
\label{null}
\ee
is satisfied, so that gravity still focuses light.

Graham \cite{grahamqfext,Graham:2005cq} 
examined the general relativistic energy conditions
required by causality.  In the neighborhood of a smooth domain wall,
given by a hyperbolic tangent, the energy density is always negative
at large enough distances.  Thus the weak energy condition is violated, as is
the null energy condition (\ref{null}).  However, when (\ref{null}) is
integrated over a complete geodesic, positivity is satisfied.  It is not clear
if this last condition, the Averaged Null Energy Condition, is always obeyed
in flat space.  Certainly it is violated in curved space, but the effects
always seem small, so that exotic effects such as time travel are prohibited.

However, as Deutsch and Candelas \cite{deutsch} showed many years ago,
in the neighborhood of a curved surface
for conformally invariant theories, $\langle T_{\mu\nu}\rangle$ diverges
as $\epsilon^{-3}$, where $\epsilon$ is the distance from the surface,
with a coefficient proportional to the sum of the principal curvatures of
the surface.  In particular they obtain the result, in the vicinity of
the surface,\index{Stress tensor!vacuum expectation value!conformal}
\begin{equation}
\langle T_{\mu\nu}\rangle\sim\epsilon^{-3}T^{(3)}_{\mu\nu}+
\epsilon^{-2}T^{(2)}_{\mu\nu}+\epsilon^{-1}T^{(1)}_{\mu\nu},
\end{equation}
and obtain explicit expressions for the coefficient tensors $T^{(3)}_{\mu\nu}$
and $T^{(2)}_{\mu\nu}$ in terms of the extrinsic curvature of the boundary.
\index{Divergences!curvature}

For example, for the case of a sphere, the leading surface divergence has the
form, for conformal fields, for $r=a+\epsilon$, $\epsilon\to0$\index{Casimir
stress!local!sphere}\index{Local Casimir effect!sphere}
\begin{equation}
\langle T_{\mu\nu}\rangle={A\over\epsilon^3}\left(\begin{array}{cccc}
2/a&\,0\,&\,0\,&0\\
0&0&0&0\\
0&0&a&0\\
0&0&0&a\sin^2\theta\end{array}\right),
\label{candelas}
\end{equation}
in spherical polar coordinates, where the constant is
$A=1/720\pi^2$ for a scalar field satisfying Dirichlet
boundary conditions, or $A=1/60\pi^2$
for the electromagnetic field satisfying perfect conductor
boundary conditions.
Note that (\ref{candelas}) is properly traceless.  The cubic divergence
in the energy density
near the surface translates into the quadratic divergence
in the energy found for a conducting ball \cite{miltonballs}.
The corresponding quadratic divergence in the stress corresponds to
the absence of the cubic divergence in $\langle T_{rr}\rangle$.

This is all completely sensible.  However, in their paper Deutsch and Candelas
\cite{deutsch} expressed a certain skepticism about the validity of the result
of \cite{mildersch} for the spherical shell case (described in part in
Sect.~\ref{Sec3.3})
where the divergences cancel.  That skepticism was reinforced in a later
paper by Candelas \cite{candelas}, who criticized the authors of
\cite{mildersch} for omitting $\delta$ function terms, and constants
in the energy.  These objections seem utterly without merit.
In a later critical paper by the same author
\cite{candelas2}, it was asserted that errors were made, rather than
a conscious removal of unphysical divergences.

Of course, surface curvature divergences are present.
As Candelas noted \cite{candelas,candelas2}, they have the form
\begin{eqnarray}
 E=E^S\int \D S+E^C\int \D S\,(\kappa_1+\kappa_2)
+E_I^C\int \D S\,(\kappa_1-\kappa_2)^2+E_{II}^C\int \D S\kappa_1\kappa_2+
\dots,
\label{divstructure}
\end{eqnarray}
where $\kappa_1$ and $\kappa_2$ are the principal curvatures of the surface.
\index{Principal curvatures} The question is
to what extent are they observable.  After all, as has been shown
in \cite{miltonbook,Milton:2002vm} and in Sect.~\ref{sec:2.4},
 we can drastically change the local structure
of the vacuum expectation value of the energy-momentum tensor
in the neighborhood of flat plates by merely
exploiting the ambiguity in the definition of that tensor, yet each
yields the same finite, observable (and observed!) energy of interaction
between the plates.  For curved boundaries, much the same is true.
{\it A priori}, we do not know which energy-momentum tensor to employ,
and the local vacuum-fluctuation energy density is to a large extent
meaningless.  It is the global energy, or the force between distinct bodies,
that has an unambiguous value.  It is the belief of the author that divergences
in the energy which go like a power of the cutoff are probably unobservable,
being subsumed in the properties of matter. Moreover, the
coefficients of the divergent terms depend on the regularization scheme.
 Logarithmic divergences, of course, are of another class \cite{bkv}.
\index{Divergences!surface|)}\index{Regularization}
\index{Casimir effect!local|)}
\index{Local effects|)}
Dramatic cancellations of these curvature terms can occur.  It might
be thought that the reason a finite result was found for the Casimir
energy of a perfectly conducting spherical
 shell \cite{Boyer:1968uf,balian,
mildersch} is that the term involving the squared difference of curvatures
in (\ref{divstructure}) is zero  only in that case.  However, 
it has been shown that at least for
the case of electromagnetism the corresponding term
is not present (or has a vanishing coefficient) for an arbitrary smooth
cavity \cite{Bernasconi}, and so the Casimir energy for a perfectly conducting
ellipsoid of revolution, for example, is finite.\footnote{The first steps 
have been made for calculating the Casimir energy for 
an ellipsoidal boundary \cite{Kitson:2005kk,kitsonromeo}, 
but only for scalar fields since the vector
Helmholtz equation is not separable in the exterior region.}
 This finiteness of the
Casimir energy (usually referred to as the vanishing of the second
heat-kernel coefficient \cite{Bordag:2001qi}) 
for an ideal smooth closed surface
was anticipated already in \cite{balian},
but contradicted by \cite{deutsch}. More specifically, although
odd curvature terms cancel inside and outside for any thin shell, it would
be anticipated that the squared-curvature term, which is present as a
surface divergence in the energy density, would be reflected as an
unremovable divergence in the energy.  For a closed surface the last term in
(\ref{divstructure})
is a topological invariant, so gives an irrelevant constant,
while no term of the type of the penultimate term can appear due to the
structure of the traced cylinder expansion \cite{Fulling:2003zx}.

\section{Casimir Forces on Spheres via $\delta$-Function Potentials}
\label{Sec3}
\index{Casimir energy for spheres}
This section is an adaptation and an extension of calculations presented
in \cite{Milton:2004vy,Milton:2004ya}.  This investigation was carried out in
response to the program of the MIT group \cite{graham,graham2,
Graham:2002fw,Graham:2003ib,Weigel:2003tp}.
They first rediscovered irremovable divergences in the Casimir energy
for a circle in $2+1$ dimensions first
discovered by Sen \cite{sen,sen2}, but then found divergences in the case of
a spherical surface, thereby casting doubt on the validity of the
Boyer calculation \cite{Boyer:1968uf}.  Some of their results, as we shall
see, are spurious, and the rest are well known \cite{bkv}.  However, their
work has been valuable in sparking new investigations of the problems of
surface energies and divergences.

We now carry out the calculation we presented in Sect.~\ref{Sec1}
in three spatial dimensions,
with a radially symmetric background \index{$\delta$-function potentials}
\be
\mathcal{L}_{\rm int}=-\frac12\frac\lambda{a^2}\delta(r-a)\phi^2(x),
\label{deltapot}
\ee
which would correspond to a Dirichlet shell in the limit $\lambda\to\infty$.
The scaling of the coupling, which here has dimensions of length,
is demanded by the requirement that the spatial integral of the potential
be independent of $a$.
The time-Fourier transformed Green's function satisfies the equation
($\kappa^2=-\omega^2$)
\be
\left[-\nabla^2+\kappa^2+\frac{\lambda}{a^2}\delta(r-a)\right]\mathcal{G}(\mathbf{r,r'})=
\delta(\mathbf{r-r'}).
\ee
We write $\mathcal{G}$ in terms of a reduced Green's function
\be
\mathcal{G}(\mathbf{r,r'})=\sum_{lm}g_l(r,r')Y_{lm}(\Omega)Y^*_{lm}(\Omega'),
\ee
where $g_l$ satisfies
\be
\left[-\frac1{r^2}\frac{\D}{\D r}r^2\frac{\D}{\D r}+
\frac{l(l+1)}{r^2}+\kappa^2
+\frac\lambda{a^2}\delta(r-a)\right]g_l(r,r')=\frac1{r^2}\delta(r-r').
\label{redgf1}
\ee
We solve this in terms of modified Bessel functions, $I_\nu(x)$, $K_\nu(x)$,
where $\nu=l+1/2$, which satisfy the Wronskian condition
\be
I'_\nu(x)K_\nu(x)-K'_\nu(x)I_\nu(x)=\frac1x.\label{wronskian}
\ee
The solution to (\ref{redgf1}) is obtained
 by requiring continuity of $g_l$ at each
singularity, at $r'$ and $a$, and the appropriate discontinuity of the
derivative.  Inside the sphere we then find ($0<r,r'<a$)
\be
 g_l(r,r')=\frac1{\kappa r r'}\left[e_l(\kappa r_>)s_l(\kappa r_<)
-\frac{\lambda}{\kappa a^2}s_l(\kappa r)s_l(\kappa r')\frac{e_l^2(\kappa a)}{1
+\frac{\lambda}{\kappa a^2}s_l(\kappa a)e_l(\kappa a)}\right].
\label{insphgf}
\ee
Here we have introduced the modified Riccati-Bessel functions,
\be
s_l(x)=\sqrt{\frac{\pi x}2}I_{l+1/2}(x),\quad
e_l(x)=\sqrt{\frac{2 x}\pi}K_{l+1/2}(x).
\ee
Note that (\ref{insphgf})
reduces to the expected Dirichlet result, vanishing as $r\to a$,
in the limit of strong coupling:
\be
\lim_{\lambda\to\infty} g_l(r,r')=\frac1{\kappa r r'}\left[e_l(\kappa r_>)
s_l(\kappa r_<)-\frac{e_l(\kappa a)}{s_l(\kappa a)}s_l(\kappa r)s_l
(\kappa r')\right].
\label{halfcasimir}
\ee
When both points are outside the sphere, $r,r'>a$, we obtain a similar
result:
\be
 g_l(r,r')=\frac1{\kappa r r'}\left[e_l(\kappa r_>)s_l(\kappa r_<)
-\frac{\lambda}{\kappa a^2}e_l(\kappa r)e_l(\kappa r')\frac{s_l^2(\kappa a)}{1
+\frac{\lambda}{\kappa a^2}s_l(\kappa a)e_l(\kappa a)}\right].
\label{outsphgf}
\ee
which similarly reduces to the expected result as $\lambda\to\infty$.

Now we want to get the radial-radial component of the stress tensor
to extract the pressure on the sphere, which is obtained by applying the
operator
\be
\partial_r\partial_{r'}-\frac12(-\partial^0\partial^{\prime0}+\bnabla
\cdot\bnabla')\to\frac12\left[\partial_r\partial_{r'}-\kappa^2
-\frac{l(l+1)}{r^2}\right]
\label{radop}
\ee to the Green's function, where in the last term we have
averaged over the surface of the sphere.  Alternatively, we
could notice that \cite{Cavero-Pelaez:15kq}
\be
\bnabla\cdot\bnabla' P_l(\cos\gamma)\bigg|_{\gamma\to0}=\frac{l(l+1)}{r^2},
\label{nablaid}
\ee
where $\gamma$ is the angle between the two directions.
In this way we find, from the discontinuity of
$\langle T_{rr}\rangle$ across the $r=a$ surface, the net stress
\be
\mathcal{S}
=-\frac{\lambda}{2\pi a^3}\sum_{l=0}^\infty  (2l+1)\int_0^\infty \D x\,
\frac{\left(e_l(x)s_l(x)\right)'-\frac{2e_l(x)s_l(x)}x}{1+\frac{\lambda a
e_l(x)s_l(x)}x}.
\label{teforce}
\ee
(Notice that there was an error in the sign of the stress, and of the scaling
of the coupling, in \cite{Milton:2004vy,Milton:2004ya}.)

The same result can be deduced by computing the total energy (\ref{casenergy}).
The free Green's function, the first term in (\ref{insphgf}) or
(\ref{outsphgf}), evidently makes no significant contribution to the energy,
for it gives a term independent of the radius of the sphere, $a$, so we
omit it.  The remaining radial integrals are simply
\alpheqn
\label{radints}
\bea
\int_0^x \D y\,s_l^2(y)&=&\frac1{2x}\left[\left(x^2+l(l+1)\right)s_l^2(x)
+xs_l(x)s_l'(x)-x^2s_l^{\prime2}(x)\right],\label{radinta}\\
\int_x^\infty \D y\,e_l^2(y)&=&
-\frac1{2x}\left[\left(x^2+l(l+1)\right)e_l^2(x)
+xe_l(x)e_l'(x)-x^2e_l^{\prime2}(x)\right].\label{radintb}
\eea
\reseteqn
Then using the Wronskian (\ref{wronskian}), we find that
the Casimir energy is
\be
E=-\frac{1}{2\pi a}\sum_{l=0}^\infty  (2l+1)\int_0^\infty \D x\,x\,
\frac{\D}{\D x}\ln\left[1+\frac{\lambda}{a} I_\nu(x)K_\nu(x)\right].
\label{teenergy}
\ee
If we differentiate with respect to $a$ we
immediately recover the force (\ref{teforce}).  This expression, upon
integration by parts, coincides with that given by Barton \cite{barton03},
and was first analyzed in detail by Scandurra \cite{Scandurra:1998xa}.
This result has also been rederived using the multiple-scattering
formalism \cite{Milton:2007wz}.
For strong coupling,
it reduces to the well-known expression for the Casimir energy
of a massless scalar field  inside and
outside a sphere upon which Dirichlet boundary conditions are imposed,
that is, that the field must vanish at $r=a$:
\be
\lim_{\lambda\to\infty}E=-\frac{1}{2\pi a}\sum_{l=0}^\infty (2l+1)\int_0^\infty
\D x\,x\,\frac{\D}{\D x}\ln\left[I_\nu(x)K_\nu(x)\right],\label{dsph}
\ee
because multiplying the argument of the logarithm by a power of $x$ is
without effect, corresponding to a contact term.  Details of the evaluation
of (\ref{dsph}) are given in \cite{Milton:2002vm}, and will be
considered in  Sect.~\ref{Sec3.3} below.  (See also
\cite{benmil,lesed1,lesed2}.)

The opposite limit is of interest here.  The expansion of the logarithm
is immediate for small $\lambda$.  The first term, of order $\lambda$,
is evidently
divergent, but irrelevant, since that may be removed by renormalization
of the tadpole graph.  In contradistinction to the claim of
\cite{graham2,Graham:2002fw,Graham:2003ib,Weigel:2003tp},
the order $\lambda^2$ term is finite,
as established in \cite{Milton:2002vm}.  That term is
\be
E^{(\lambda^2)}=\frac{\lambda^2}{4\pi a^3}
\sum_{l=0}^\infty(2l+1)\int_0^\infty \D x\,x
\frac{\D}{\D x}[I_{l+1/2}(x)K_{l+1/2}(x)]^2.\label{og}
\ee
The sum on $l$ can be carried out using a trick due to Klich \cite{klich}:
The sum rule
\be
\sum_{l=0}^\infty (2l+1)e_l(x)s_l(y)P_l(\cos\theta)=\frac{xy}\rho \E^{-\rho},
\ee
where $\rho=\sqrt{x^2+y^2-2xy\cos\theta}$, is squared, and then integrated
over $\theta$, according to
\be
\int_{-1}^1 \D(\cos\theta)
P_l(\cos\theta)P_{l'}(\cos\theta)=\delta_{ll'}\frac2{2l+1}.
\ee
In this way we learn that
\be
\sum_{l=0}^\infty (2l+1)e_l^2(x)s_l^2(x)=\frac{x^2}2\int_0^{4x}\frac{\D w}w
\E^{-w}.
\ee
Although this integral is divergent, because we did not integrate by parts
in (\ref{og}), that divergence does not contribute:
\be
E^{(\lambda^2)}=\frac{\lambda^2}{4\pi a^3}\int_0^\infty \D x\,
\frac12 x \,\frac{\D}{\D x}\int_0^{4x}
\frac{\D w}w \E^{-w}=\frac{\lambda^2}{32\pi a^3},
\label{4.25}
\ee
which is exactly the result (4.25) of \cite{Milton:2002vm}.

However, before we wax too euphoric, we recognize that the order $\lambda^3$
term
appears logarithmically divergent, just as \cite{Graham:2003ib} and
\cite{Weigel:2003tp}
claim.  This does not signal a breakdown in perturbation
theory.  Suppose we
subtract off the two leading terms,
\bea
 E&=&-\frac1{2\pi a}\sum_{l=0}^\infty(2l+1)\int_0^\infty \D x\,x\,\frac{\D}
{\D x}
\left[\ln\left(1+\frac\lambda{a} I_\nu K_\nu\right)-\frac\lambda{a} a I_\nu K_\nu+
\frac{\lambda^2}{2a^2}(I_\nu K_\nu)^2
\right]\nn\\
&&\quad\mbox{}+\frac{\lambda^2}{32\pi a^3}.\label{fulle}
\eea
To study the behavior of the sum for large values of $l$, we can use the
uniform asymptotic expansion (Debye expansion), for $\nu\to\infty$,
\label{Uniform asymptotic expansion}
\bea
I_\nu(x)&\sim&\sqrt{\frac{t}{2\pi \nu}}\E^{\nu\eta}\left(1+\sum_k\frac{u_k(t)}{\nu^k}
\right),\nn\\
K_\nu(x)&\sim&\sqrt{\frac{\pi t}{2 \nu}}\E^{-\nu\eta}\left(1+\sum_k(-1)^k
\frac{u_k(t)}{\nu^k}\right),\label{uae}
\eea
where 
\be
x=mz, \quad t=1/\sqrt{1+z^2}, \quad \eta(z) = \sqrt{1 + z^2}
+ \ln \left[\frac{z}{1 + \sqrt{1+z^2}}
\right],\quad \frac{\D\eta}{\D z}=\frac1{zt}.\label{zteta}
\ee
The polynomials in $t$ appearing in (\ref{uae}) are generated by
\bea
u_0(t)&=&1,\quad
u_k(t)=\frac12t^2(1-t^2)u'_{k-1}(t)+\frac18\int_0^t \D s(1-5s^2)u_{k-1}(s).
\eea
 We now insert these expansions into (\ref{fulle}) and expand
not in $\lambda$ but in $\nu$; the leading term is
\be
E^{(\lambda^3)}\sim
\frac{\lambda^3}{24\pi a^4}\sum_{l=0}^\infty\frac1\nu\int_0^\infty
\frac{\D z}{(1+z^2)^{3/2}}=\frac{\lambda^3}{24\pi a^4}\zeta(1).
\ee
Although the frequency integral is finite, the angular momentum sum is
divergent.  The appearance here of the divergent $\zeta(1)$ seems to
signal an insuperable barrier to extraction of a finite Casimir energy
for finite $\lambda$.  The situation is different in the limit
$\lambda\to\infty$ ---See Sect.~\ref{Sec3.3}.

\index{Divergence in total energy}
This divergence has been known for many years, and was first calculated
explicitly in 1998 by Bordag et al.~\cite{bkv}, where the second heat kernel
coefficient gave an equivalent result,
\be
E\sim \frac{\lambda^3}{48\pi a^4}\frac1s,\quad s\to0.\label{secondhk}
\ee
A possible way of dealing with this divergence was advocated in
\cite{Scandurra:1998xa}. More recently, Bordag and Vassilevich
\cite{Bordag:2004rx}  have
reanalyzed such problems from the heat kernel approach.  They show that this
$\mathrm{O}(\lambda^3)$ divergence corresponds to a surface 
tension counterterm,
an idea proposed by me in 1980 \cite{miltonbag,miltonfermion} in connection
with the zero-point energy contribution to the bag model.  Such a surface
term corresponds to $\lambda$ fixed, which then necessarily implies
a divergence of order $\lambda^3$.  Bordag argues that it is perfectly
appropriate to insert a surface tension counterterm so that this divergence
may be rendered finite by renormalization.\index{Renormalization}

\subsection{TM Spherical Potential}
\label{Sec3.2}
\index{Transverse magnetic modes}
Of course, the scalar model considered in the previous subsection is merely
a toy model, and something analogous to electrodynamics is of far more
physical relevance.  There are good reasons for believing that cancellations
occur in general between TE (Dirichlet) and TM (Robin) modes.  Certainly
they do occur in the classic Boyer energy of a perfectly conducting spherical
shell \cite{Boyer:1968uf, balian, mildersch}, and the indications
are that such cancellations occur even with imperfect boundary conditions
\cite{barton03}.  Following the latter reference, let us consider the
potential
\be
\mathcal{L}_{\rm int}=\frac12\lambda\frac1r\frac\partial{\partial r}
\delta(r-a)\phi^2(x).
\label{tmlag}
\ee
Here $\lambda$ again has dimensions of length.
In the limit $\lambda\to\infty$ this corresponds to TM boundary conditions.
The reduced Green's function is thus taken to satisfy
\be
\left[-\frac1{r^2}\frac{\partial}{\partial r}r^2\frac{\partial}{\partial r}+\frac{l(l+1)}{r^2}
+\kappa^2-\frac{\lambda}r\frac\partial{\partial r}\delta(r-a)\right]g_l(r,r')
=\frac1{r^2}\delta(r-r').
\ee
At $r=r'$ we have the usual boundary conditions, that $g_l$ be continuous, but
that its derivative be discontinuous,
\be
r^2\frac{\partial}{\partial r}g_l\bigg|_{r=r'-}^{r=r'+}=-1,
\ee
while at the surface of the sphere the derivative is continuous,
\alpheqn
\be
\frac\partial{\partial r}rg_l\bigg|_{r=a-}^{r=a+}=0,
\label{dcont}
\ee
while the function is discontinuous,
\be
g_l\bigg|_{r=a-}^{r=a+}=-\frac{\lambda}a 
\frac\partial{\partial r}rg_l\bigg|_{r=a}.
\label{fndiscont}
\ee
\reseteqn
Equations (\ref{dcont}) and (\ref{fndiscont}) are the analogues of the
boundary conditions (\ref{tmbc1}), (\ref{tmbc2}) treated in 
Sect.~\ref{sec:tm11}.

 It is then easy to find the Green's function.  When both points are
inside the sphere,
\alpheqn
\be
r,r'<a:\quad g_l(r,r')=\frac1{\kappa rr'}\left[s_l(\kappa r_<)e_l(\kappa r_>)
-\frac{\lambda\kappa[e_l'(\kappa a)]^2s_l(\kappa r)s_l(\kappa r')}
{1+\lambda\kappa e_l'(\kappa a)s_l'(\kappa a)}\right],
\ee
and when both points are outside the sphere,
\be
r,r'>a:\quad g_l(r,r')=\frac1{\kappa rr'}\left[s_l(\kappa r_<)e_l(\kappa r_>)
-\frac{\lambda\kappa[s_l'(\kappa a)]^2e_l(\kappa r)e_l(\kappa r')}
{1+\lambda\kappa e_l'(\kappa a)s_l'(\kappa a)}\right].
\ee
\reseteqn
It is immediate that these supply the appropriate Robin boundary conditions
in the $\lambda\to\infty$ limit:
\be
\lim_{\lambda\to0}\frac\partial{\partial r}rg_l\bigg|_{r=a}=0.
\ee

The Casimir energy may be readily obtained from (\ref{casenergy}),
and we find, using the integrals (\ref{radinta}), (\ref{radintb})
\be
E=-\frac1{2\pi a}\sum_{l=0}^\infty (2l+1)\int_0^\infty \D x\,x\frac{\D}
{\D x}
\ln\left[1+\frac{\lambda}a x e_l'(x)s_l'(x)\right].
\label{tmenergy}
\ee
The stress may be obtained from this by applying $-\partial/\partial a$,
and regarding $\lambda$ as constant,
or directly, from the Green's function by applying the operator,
\be
t_{rr} =\frac1{2\I}\left[\nabla_r\nabla_{r'}-\kappa^2-\frac
{l(l+1)}{r^2}\right]g_l\bigg|_{r'=r},
\ee
which is the same as that in (\ref{radop}), except that
\be
\nabla_r=\frac1r\partial_r r,
\ee
appropriate to TM boundary conditions (see \cite{mildim}, for example).
Either way, the total stress on the sphere is
\be
\mathcal{S}=
-\frac\lambda{2\pi a^3}\sum_{l=0}^\infty(2l+1)\int_0^\infty dx\,x^2
\frac{[e_l'(x)
s_l'(x)]'}{1+\frac\lambda{a} x e_l'(x)s_l'(x)}.
\ee
The result for the energy (\ref{tmenergy}) is similar, but not identical, to
that given by Barton \cite{barton03}.

Suppose we now combine the TE and TM Casimir energies, (\ref{teenergy})
and (\ref{tmenergy}):
\be
 E^{\rm TE}+E^{\rm TM}=-\frac1{2\pi a}\sum_{l=0}^\infty (2l+1)\int_0^\infty
\D x\,x\frac \D{\D x}
\ln\left[\left(1+\frac{\lambda}a\frac{e_ls_l}x\right)\left(1+\frac\lambda{a} x
e_l's_l'\right)\right].
\label{combenergy}
\ee
In the limit $\lambda\to\infty$ this reduces to the familiar expression
for the perfectly conducting spherical shell \cite{mildersch}:
\be
\lim_{\lambda\to\infty}E=-\frac1{2\pi a}\sum_{l=1}^\infty(2l+1)\int_0^\infty
\D x\,x\left(\frac{e_l'}{e_l}+\frac{e_l''}{e_l'}+\frac{s_l'}{s_l}
+\frac{s_l''}{s_l'}\right).
\label{spherece}
\ee
Here we have, as appropriate to the electrodynamic situation, omitted the
$l=0$ mode. This expression
 yields a finite Casimir energy, as we will see in Sect.~\ref{Sec3.3}.
What about finite $\lambda$?  In general,
it appears that there is no chance that the divergence found in the previous
section in order $\lambda^3$ can be cancelled.  But suppose the coupling
for the TE and TM modes are different.  If $\lambda^{\rm TE}\lambda^{\rm
TM}=4 a^2$, a cancellation appears possible, as discussed in 
\cite{Milton:2004ya}.

\subsection{Evaluation of Casimir Energy for
a Dirichlet Spherical Shell}
\label{Sec3.3} \index{Casimir energy for perfectly conducting sphere}
In this subsection  we will evaluate the
above expression (\ref{dsph})
 for the Casimir energy for a massless scalar in three space
dimensions, with a spherical boundary on which the field vanishes.
This corresponds to the TE modes for the electrodynamic situation first
solved by Boyer \cite{Boyer:1968uf,balian,mildersch}.   The purpose of this
section (adapted from \cite{Milton:2002vm,Milton:2004ya}) is to
emphasize anew that, contrary to the implication of \cite{graham2,
Graham:2002fw,Graham:2003ib,Weigel:2003tp},
the corresponding Casimir energy is also finite for this configuration.

 The general calculation in $D$ spatial
dimensions was given in \cite{benmil}; the pressure is
given by the formula \index{Dimensional continuation}
\be
 P=-\sum_{l=0}^\infty\frac{(2l+D-2)\Gamma(l+D-2)}{l!2^D\pi^{(D+1)/2}
\Gamma(\frac{D-1}2)a^{D+1}}\int_0^\infty \D x\,x\frac{\D}{\D x}\ln\left[I_\nu(x)
K_\nu(x)x^{2-D}\right].\label{3.1}\ee
Here $\nu=l-1+D/2$.  For $D=3$ this expression reduces to
\be
P=-\frac1{8\pi^2a^4}\sum_{l=0}^\infty(2l+1)\int_0^\infty\D x\,x
\frac{\D}
{\D x}\ln\left[I_{l+1/2}(x)K_{l+1/2}(x)/x\right].
\label{fsphere}
\ee
This precisely corresponds to the strong limit $\lambda\to\infty$ given
in (\ref{dsph}), if we recall the comment made about contact terms there.
In \cite{benmil} we evaluated  expression (\ref{3.1})
by continuing in $D$ from
a region where both the sum and integrals existed.  In that way, a completely
finite result was found for all positive $D$ not equal to an even integer.

Here we will adopt a perhaps more physical approach, that of allowing the
time-coordinates in the underlying Green's function to approach each other,
temporal point-splitting,
as described in \cite{mildersch}.  That is, we recognize that the $x$
integration above is actually a (dimensionless) imaginary
frequency integral, and therefore we should replace
\be
\int_0^\infty \D x\,f(x)=\frac12\int_{-\infty}^\infty \D y\,\E^{\I
y\delta}f(|y|),
\label{timesplit}
\ee
where at the end we are to take $\delta\to0$.  Immediately, we can
replace the $x^{-1}$ inside the logarithm in (\ref{fsphere})
 by $x$, which makes the integrals
converge, because the difference is proportional to a $\delta$ function in
the time separation, a contact term without physical significance.

To proceed, we use the uniform asymptotic expansions for the modified
Bessel functions, (\ref{uae}).  This
is an expansion in inverse powers of $\nu=l+1/2$, low terms in which turn
out to be remarkably accurate even for modest $l$.  The leading terms
in this expansion are, using (\ref{uae}),\index{Uniform asymptotic expansion}
\be
\ln\left[x I_{l+1/2}(x)K_{l+1/2}(x)\right]
\sim\ln\frac{zt}2+\frac1{\nu^2}g(t)+\frac1{\nu^4}
h(t)+\dots,
\label{uae2}
\ee
\alpheqn
\bea
g(t)&=&\frac18(t^2-6t^4+5t^6),\\
h(t)&=&\frac1{64}(13t^4-284t^6+1062t^8-1356t^{10}+565t^{12}).
\eea
\reseteqn
The leading term in the pressure is therefore
\bea
 P_0=
-\frac1{8\pi^2a^4}\sum_{l=0}^\infty(2l+1)\nu\int_0^\infty \D z\, t^2
=-\frac1{8\pi a^4}\sum_{l=0}^\infty\nu^2=\frac3{32\pi a^4}\zeta(-2)=0,\nn\\
\eea
where in the last step we have used the formal zeta function
evaluation\footnote{Note that the corresponding TE 
contribution
the electromagnetic Casimir pressure would not be zero, for there the sum
starts from $l=1$.}
\be
\sum_{l=0}^\infty \nu^{-s}=(2^s-1)\zeta(s).\label{zetafn}
\ee
 Here the rigorous way to argue is to recall the
presence of the point-splitting factor $\E^{\I \nu z\delta}$ and to carry out
the sum on $l$ using
\be
\sum_{l=0}^\infty \E^{\I\nu z\delta}=-\frac1{2\I}\frac1{\sin z\delta/2},
\label{lsum}
\ee
so
\bea
\sum_{l=0}^\infty \nu^2\E^{\I\nu z\delta}=-\frac{\D^2}{\D(z\delta)^2}
\frac{\I}
{2\sin z\delta/2}
=\frac{\I}8\left(-\frac2{\sin^3z\delta/2}+\frac1{\sin z\delta/2}\right).
\eea
Then $P_0$ is given by the divergent expression
\be
P_0=\frac{\I}{4\pi^2 a^4\delta^3}\int_{-\infty}^\infty \frac{\D z}{z^3}
\frac1{1+z^2},
\ee
which we argue is zero because the integrand is odd, as justified by averaging
over contours passing above and below the pole at $z=0$.

The next term in the uniform asymptotic expansion (\ref{uae2}),
that involving $g$,
likewise gives zero pressure, as intimated by (\ref{zetafn}),
which vanishes at $s=0$.  The same conclusion follows from point splitting,
using (\ref{lsum}) and arguing that the resulting integrand
$\sim z^2t^3 g'(t)/z\delta$ is odd in $z$.
  Again, this cancellation does not occur in the electromagnetic case
because there the sum starts at $l=1$.

So here the leading term which survives is that of order $\nu^{-4}$
in (\ref{uae2}),
namely
\be
P_2=\frac1{4\pi^2a^4}\sum_{l=0}^\infty \frac1{\nu^2}\int_0^\infty
\D z \,h(t),
\ee
where we have now dropped the point-splitting factor because this expression
is completely convergent.  The integral over $z$ is
\be
\int_0^\infty \D z \, h(t)=\frac{35\pi}{32768}
\ee and the sum over $l$ is $3\zeta(2)=\pi^2/2$, so the leading contribution
to the stress on the sphere is
\be
{\cal S}_2=4\pi a^2P_2=\frac{35\pi^2}{65536a^2}=\frac{0.00527094}{a^2}.
\ee
Numerically this is a terrible approximation.

What we must do now is return to the full expression and add and subtract
the leading asymptotic terms.  This gives
\be
{\cal S}={\cal S}_2-\frac1{2\pi a^2}\sum_{l=0}^\infty(2l+1)R_l,
\ee
where
\be
R_l=Q_l+\int_0^\infty \D x\left[\ln zt+\frac1{\nu^2}g(t)+\frac1{\nu^4}h(t)\right],
\label{remainder}
\ee
where the integral
\be
Q_l=-\int_0^\infty \D x\ln[2xI_\nu(x)K_{\nu}(x)]
\ee
was given the asymptotic form  in \cite{benmil,miltonbook} ($l\gg1$):
\bea
 Q_l&\sim&\frac{\nu\pi}2+\frac\pi{128\nu}-\frac{35\pi}{32768\nu^3}
+\frac{565\pi}{1048577\nu^5}
-\frac{1208767\pi}{2147483648\nu^7}\nn\\
&&\qquad\mbox{}+\frac{138008357\pi}{137438953472\nu^9}+\dots.
\label{ql}
\eea\index{Uniform asymptotic expansion}
The first two terms in (\ref{ql}) cancel the second and third terms in
(\ref{remainder}), of course.
The third term in (\ref{ql}) corresponds to $h(t)$, so the last three terms
displayed in (\ref{ql}) give the asymptotic behavior of the remainder,
which we call $w(\nu)$.  Then we have, approximately,
\be
{\cal S}\approx {\cal  S}_2-\frac1{\pi a^2}\sum_{l=0}^n\nu R_l-\frac1{\pi a^2}
\sum_{l=n+1}^\infty \nu w(\nu).
\ee
For $n=1$ this gives ${\cal S}\approx0.002\,852\,78/a^2$, and for larger $n$
this rapidly approaches the value first  given in \cite{benmil},
and rederived in \cite{lesed1,lesed2,lesedflux}
\be
{\cal S}^{\rm TE}=0.002817/a^2,
\ee
a value much smaller than the famous electromagnetic result \cite{Boyer:1968uf,
davis,mildersch,balian},
\be
{\cal S}^{\rm EM}=\frac{0.04618}{a^2},
\label{boyerresult}
\ee
because of the cancellation of the leading terms noted above.
Indeed, the TM contribution was calculated separately in \cite{mildim},
with the result
\be
\mathcal{S}^{\rm TM}=-0.02204\frac1{a^2},
\ee
and then subtracting the $l=0$ modes from both contributions we obtain
(\ref{boyerresult}),
\be
{\cal S}^{\rm EM}=\mathcal{S}^{\rm TE}+\mathcal{S}^{\rm TM}+\frac{\pi}{48a^2}
=\frac{0.0462}{a^2}.
\ee


\subsection{Surface Divergences in the Energy Density}\label{Sec:surfdiv}
\index{Surface divergences}
The following discussion is based on \cite{Cavero-Pelaez:15kq}.
Using (\ref{nablaid}), we immediately find
the following expression for the energy density
inside or outside the sphere:
\bea
\langle T^{00}\rangle &=&\int_0^\infty \frac{\D\kappa}{2\pi}\sum_{l=0}^\infty
\frac{2l+1}{4\pi}\left\{\left[-\kappa^2+\partial_r\partial_{r'}+
\frac{l(l+1)}{r^2}
\right]g_{l}(r,r')\bigg|_{r'=r}\right.\nn\\
&&\qquad\qquad\mbox{}-\left.2\xi\frac1{r^2}\frac{\partial}{\partial r}
r^2\frac{\partial}{\partial r}g_l(r,r)\right\},\label{t00}
\eea
where $\xi$ is the conformal parameter as seen in (\ref{stconformal}).
To find the energy density in either region
we insert the appropriate Green's functions (\ref{insphgf}) or
(\ref{outsphgf}), but delete the free part,
\bea
 g^0_l&=&\frac1{\kappa rr'}s_l(\kappa r_<)e_l(\kappa r_>),
\eea
which corresponds to the {\em bulk energy\/} which would be
present if either medium filled all of space, leaving us with for $r>a$
\bea
u(r)&=&-(1-4\xi)\int_0^\infty \frac{\D\kappa}{2\pi}\sum_{l=0}^\infty 
\frac{2l+1}{4\pi}\frac{\frac\lambda{\kappa a^2}s_l^2(\kappa a)}{1+\frac{\lambda}{\kappa a^2}
e_l(\kappa a)s_l(\kappa a)}
\bigg\{\frac{e_l^2(\kappa r)}{\kappa r^2}
\left[-\kappa^2\frac{1+4\xi}{1-4\xi}\right.\nn\\
&&\quad\mbox{}+\left.\frac{l(l+1)}{r^2}+\frac1{r^2}
\right]-\frac2{r^3}e_l(\kappa r)e_l'(\kappa r)+\frac\kappa
{r^2}e_l^{\prime2}(\kappa r)\bigg\}.\label{uout}
\eea
Inside the shell, $r<a$, the energy is given by a similar expression
obtained from (\ref{uout}) by interchanging $e_l$ and $s_l$. 

We want to examine the singularity structure as $r\to a$
from the outside.   For this purpose
we use the leading uniform asymptotic expansion, $l\to\infty$,
obtained from (\ref{uae})\index{Uniform asymptotic expansion}
\bea
 e_l(x)&\sim& \sqrt{zt}\,e^{-\nu \eta},\quad
s_l(x)\sim\frac12\sqrt{zt}\,e^{\nu \eta},\nonumber\\
e_l'(x)&\sim&-\frac1{\sqrt{zt}}\,e^{-\nu\eta},\quad s'_l(x)\sim
\frac12\frac1{\sqrt{zt}}\,e^{\nu \eta},\label{uaex}
\eea
where $\nu=l+1/2$, and $z$, $t$, and $\eta$ are given in (\ref{zteta}).
The coefficient of $e_l(\kappa r)e_l(\kappa r')$ 
occurring in the $\delta$-function
potential Green's function (\ref{outsphgf}),
in strong and weak coupling, becomes
\alpheqn
\bea
\frac{\lambda}a \to\infty: &\to&
\frac{s_l(\kappa a)}{e_l(\kappa a)},\\
\frac{\lambda}a\to0:\quad  &\to&
\frac\lambda{\kappa a^2}s_l^2(\kappa a).
\eea
\reseteqn

In either case, we carry out the asymptotic sum over angular momentum using
(\ref{uaex}) and the analytic continuation of (\ref{lsum})
\bea
\sum_{l=0}^\infty e^{-\nu\chi}=\frac1{2\sinh\frac\chi2}.\label{sumoverl}
\eea
Here ($r\approx a$)
\be
\chi=2\left[\eta(z)-\eta\left(z\frac{a}r\right)\right]\approx
2z\frac{\D\eta}{\D z}(z)\frac{r-a}r=\frac2t\frac{r-a}r.
\ee
The remaining integrals over $z$ are elementary, and in this
way we find that the leading divergences
in the energy density are as $r\to a+$,
\alpheqn
\bea
\frac\lambda{a}\to\infty:\quad u&\sim&
-\frac1{16\pi^2}\frac{1-6\xi}{(r-a)^4},\label{dsphere1}\\
\frac\lambda{a}\to0:\quad  u^{(n)}&\sim&
\left(-\frac\lambda{a}\right)^n\frac{\Gamma(4-n)}{96\pi^2a^4}(1-6\xi)
\left(\frac{a}{r-a}\right)^{4-n},\quad n<4,\label{surfdivn}
\eea
\reseteqn
where the latter is  the leading divergence in order $n$.
These results clearly seem to demonstrate the virtue of the conformal
value of $\xi=1/6$; but see below.
(The  value for the Dirichlet sphere (\ref{dsphere1}) first appeared
in \cite{deutsch}; it more
recently was rederived in \cite{Schwartz-Perlov:2005ds},
where, however, the subdominant term, the leading
term if $\xi=1/6$, namely (\ref{sdsc}), was not calculated.
Of course, this result is the same as the surface divergence
encountered for parallel Dirichlet plates \cite{Milton:2002vm,miltonbook}.)
The perturbative divergence for $n=1$ in (\ref{surfdivn}) is exactly
that found for a plate---see (\ref{pertsd}).

Thus, for $\xi=1/6$ we must keep subleading terms.
This includes keeping the subdominant term in $\chi$,\footnote{Note
there is a sign error in (4.8) of \cite{Cavero-Pelaez:15kq}.}
\be
\chi\approx\frac2t\frac{r-a}r+t\left(\frac{r-a}r\right)^2,\label{chiexp}
\ee
the distinction between $t(z)$ and $\tilde t=t(\tilde z=za/r)$,
\be
\tilde z\tilde t\approx zt-t^3z\frac{r-a}r,\label{tztd}
\ee
as well as the next term in the uniform asymptotic expansion of the
Bessel functions (\ref{uae}).
Including all this, it is straightforward to recover the
well-known result (\ref{candelas}) \cite{deutsch}
for strong coupling (Dirichlet boundary conditions):
\be
 \frac\lambda{a}\to\infty:\quad  u\sim
\frac1{360\pi^2}\frac1{a(r-a)^3},\label{sdsc}
\ee
Following the same process for
weak coupling, we find that the
 leading divergence in order $n$, $1\le n<3$, is ($r\to  a\pm$)
\be
\lambda\to 0:\quad u^{(n)}\sim\left(\frac\lambda{a^2}\right)^{n}
\frac1{1440\pi^2}\frac1{a(a-r)^{3-n}}(n-1)(n+2)\Gamma(3-n).\label{sdwc}
\ee
Note that the subleading $O(\lambda)$ term again vanishes.
Both Eqs.~(\ref{sdsc}) and (\ref{sdwc}) apply
for the conformal value $\xi=1/6$.

\subsection{Total Energy and Renormalization}
\index{Renormalization}

As discussed in \cite{Cavero-Pelaez:15kq}  
we may consider the potential, in the spirit of (\ref{bkgdpot}),
\alpheqn\label{smoothpot}
\be \mathcal{L}_{\rm int}=-\frac\lambda{2a^2}\phi^2\sigma(r),
\ee
where
\be
\sigma(r)=\left\{\begin{array}{cc}
0,&r<a_-,\\
h,&a_-<r<a_+,\\
0,&a_+<r.\end{array}\right.
\ee
\reseteqn
Here $a_\pm=a\pm\delta/2$, and we set $h\delta=1$.
That is, we have expanded the $\delta$-function shell so that
it has finite thickness.

In particular,  the integrated local energy density
inside, outside, and within the shell is $E_{\rm in}$, $E_{\rm out}$, and
$E_{\rm sh}$, respectively.  The total energy of a given region is the sum of
the integrated local energy and the surface energy (\ref{es1}) bounding that 
region ($\xi=1/6$):\index{Surface energy}
\alpheqn
\bea
\tilde E_{\rm in}&=&E_{\rm in}+\hat E_-,\\
\tilde E_{\rm out}&=&E_{\rm out}+\hat E_+,\\
\tilde E_{\rm sh}&=&E_{\rm sh}+\hat E'_++\hat E'_-,
\eea
\reseteqn
where $\hat E_\pm$ is the outside (inside) surface energy on the
surface at $r=a_\pm$, while $\hat E_\pm'$ is the inside (outside) surface
energy on the same surfaces.  $E_{\rm in}$, $E_{\rm out}$, and $E_{\rm sh}$
represent $\int (\D\mathbf{r}) \langle T^{00}\rangle$ in each region.
 Because for a nonsingular potential the surface
energies cancel across each boundary,
\be
\hat E_++\hat E_+'=0,\quad\hat E_-+\hat E_-'=0,
\ee
the total energy is
\be
E=\tilde E_{\rm in}+\tilde E_{\rm out}+\tilde E_{\rm sh}
=E_{\rm in}+E_{\rm out}+E_{\rm sh}.
\ee
In the singular thin shell limit, the integrated local shell energy
is the total surface energy of a thin Dirichlet shell:
\be
E_{\rm sh}=\hat E_++\hat E_-\ne 0.
\ee
See the remark at the end of Sect.~\ref{sec:2.4}.
This shell energy, for the conformally coupled
theory, is finite in second order in the
coupling (in at least two plausible regularization schemes),
but diverges in third order.  We showed in \cite{Cavero-Pelaez:15kq} 
that the latter precisely
corresponds to the known divergence of the total energy in this order.
Thus we have established the suspected correspondence between surface
divergences and divergences in the total energy, which has nothing to
do with divergences in the local energy density as the surface is approached.
This precise correspondence should enable us to absorb such
global divergences in
a renormalization of the surface energy, and should lead to further advances
of our understanding of quantum vacuum effects.
We will elaborate on this point in the following.

\section{Semitransparent Cylinder}
\label{sec:gf}\index{Casimir effect for cylinder}
This section is based on \cite{CaveroPelaez:2006rt}.
We consider a massless scalar field $\phi$ in a $\delta$-cylinder background,
\index{$\delta$-function potential}
\be
\mathcal{L}_{\rm int}=-\frac\lambda{2a}\delta(r-a)\phi^2,
\ee
$a$ being the radius of the ``semitransparent'' cylinder. The
massive case was earlier considered by Scandurra \cite{Scandurra:2000qz}.
We will continue to  assume that
the dimensionless coupling $\lambda>0$ to avoid the appearance of negative eigenfrequencies.
The time-Fourier transform of the Green's function
satisfies
\be
\left[-\nabla^2-\omega^2+\lambda\delta(r-a)\right]\mathcal{G}
(\mathbf{r,r'})=\delta(\mathbf{r-r'}).
\ee
Adopting cylindrical coordinates, we write
\be
\mathcal{G}(\mathbf{r,r'})=\int\frac{\D k}{2\pi}\E^{\I k(z-z')}
\sum_{m=-\infty}^\infty
\frac1{2\pi}e^{\I m(\varphi-\varphi')}g_m(r,r';k),
\ee
where the reduced Green's function satisfies
\be
\left[-\frac1r\frac{\partial}{\partial r}r\frac{\partial}{\partial r}+\kappa^2+\frac{m^2}{r^2}
+\frac{\lambda}a
\delta(r-a)\right]g_m(r,r';k)=\frac1r\delta(r-r'),\label{rgfeqn}
\ee
where $\kappa^2=k^2-\omega^2$.  Let us immediately make a Euclidean rotation,
\be
\omega\to \I\zeta,
\ee
where $\zeta$ is real, so $\kappa$ is likewise always real.  Apart from
the $\delta$ functions, this is the modified Bessel equation.

Because of the Wronskian (\ref{wronskian}) satisfied by the modified Bessel functions,
we have the general solution to (\ref{rgfeqn}) as long as $r\ne a$ to be
\be
g_m(r,r';k)=I_m(\kappa r_<)K_m(\kappa r_>)+A(r')I_m(\kappa r)+B(r')
K_m(\kappa r),
\ee
where $A$ and $B$ are arbitrary functions of $r'$.
Now we incorporate the effect of the $\delta$ function at $r=a$
in (\ref{rgfeqn}).
It implies that $g_m$ must be continuous at $r=a$, while it has a discontinuous
derivative,
\be
\frac{\partial}{\partial r}g_m(r,r';k)\bigg|_{r=a-}^{r=a+}=\frac{\lambda}a g_m(a,r';k),
\ee
from which we rather immediately deduce the form of the Green's function
inside and outside the cylinder:
\alpheqn
\bea
r,r'<a:\quad g_m(r,r';k)&=&I_m(\kappa r_<)K_m(\kappa r_>)\nonumber\\
&&\quad\mbox{}-\frac{\lambda 
K_m^2(\kappa a)}{1+\lambda  I_m(\kappa a)K_m(\kappa a)}I_m(\kappa r)
I_m(\kappa r'),\label{gin2}
\\
r,r'>a:\quad g_m(r,r';k)&=&I_m(\kappa r_<)K_m(\kappa r_>)\nonumber\\
&&\quad\mbox{}-\frac{\lambda 
I_m^2(\kappa a)}{1+\lambda  I_m(\kappa a)K_m(\kappa a)}K_m(\kappa r)
K_m(\kappa r').\label{gout2}
\eea
\reseteqn
Notice that in the limit $\lambda\to\infty$ we recover the Dirichlet cylinder
result, that is, that $g_m$ vanishes at $r=a$.

\subsection{Cylinder Pressure and Energy}
\label{sec:pressure}

The easiest way to calculate the total energy is to compute the pressure on the
cylindrical walls due to the quantum fluctuations in the field.  This may be
computed, at the one-loop level, from the vacuum expectation value of the
stress tensor,\index{Conformal stress tensor}
\be
\langle T^{\mu\nu}\rangle
=\left(\partial^\mu\partial^{\prime\nu}-\frac12 g^{\mu\nu}
\partial^\lambda\partial'_\lambda\right)\frac1\I G(x,x')\bigg|_{x=x'}-\xi
(\partial^\mu\partial^\nu-g^{\mu\nu}\partial^2)\frac1\I G(x,x),\label{st2}
\ee
which we have written in a Cartesian coordinate system.
Here we have again 
included the conformal parameter $\xi$, which is equal to 1/6
for the stress tensor that makes conformal invariance manifest.  The conformal
term does not contribute to the radial-radial component of the stress tensor,
however, because then only transverse and time derivatives act on $G(x,x)$,
which depends only on $r$.  The discontinuity of the expectation value of the
radial-radial
component of the stress tensor is the pressure of the cylindrical wall:
\bea
P&=&\langle T_{rr}\rangle_{\rm in}-\langle T_{rr}\rangle_{\rm out}\nonumber\\
&=&-\frac1{16\pi^3}\sum_{m=-\infty}^\infty \int_{-\infty}^\infty \D k
\int_{-\infty}^\infty 
\D\zeta\frac{\lambda \kappa^2}{1+\lambda  I_m(\kappa a)K_m(\kappa a)}
\nonumber\\
&&\qquad\times
\left[K_m^2(\kappa a)I_m^{\prime 2}(\kappa a)-I_m^2(\kappa a)K_m^{\prime 2}
(\kappa a)\right]\nonumber\\
&=&-\frac1{16\pi^3}\sum_{m=-\infty}^\infty\int_{-\infty}^\infty \D k
\int_{-\infty}^\infty \D\zeta\frac\kappa{a}\frac{\D}
{\D\kappa a}\ln\left[1+\lambda I_m(\kappa a)
K_m(\kappa a)\right],
\eea
where we have again used the Wronskian (\ref{wronskian}) .  Regarding $ka$ and
$\zeta a$ as the two Cartesian components of a two-dimensional vector, with
magnitude $x\equiv\kappa a=\sqrt{k^2a^2+\zeta^2 a^2}$, we get the stress on the
cylinder per unit length to be
\be
\mathcal{S}=2\pi a P=-\frac1{4\pi a^3}\int_0^\infty \D x\,x^2
\sum_{m=-\infty}^\infty
\frac{\D}{\D x}\ln\left[1+\lambda  I_m(x)K_m(x) \right],
\ee
which possesses the expected Dirichlet limit as $\lambda\to\infty$.
The corresponding expression for the total Casimir energy per unit length
follows by integrating
\be
\mathcal{S}=-\frac\partial{\partial a}\mathcal{E},
\ee
that is,
\be
\mathcal{E}=-\frac1{8\pi a^2}\int_0^\infty \D x\,x^2\sum_{m=-\infty}^\infty
\frac{\D}{\D x}\ln\left[1+\lambda I_m(x)K_m(x) \right].\label{energy}
\ee
This expression, the analog of (\ref{teenergy}) for the spherical case,
 is, of course, completely formal, and will be regulated in
various ways, for example, with an analytic or exponential
regulator  as we will see in the following, or by using
zeta-function regularization \cite{CaveroPelaez:2006rt}.

Alternatively, we may compute the energy directly from the general
formula (\ref{casenergy}).
To evaluate (\ref{casenergy}) in this case, we use the
standard indefinite integrals over squared Bessel functions.
When we insert the above construction of the Green's function
(\ref{gin2}) and (\ref{gout2}), and perform
the integrals over the regions interior and exterior to the
cylinder we obtain (\ref{energy}) immediately.

\subsection{Weak-coupling Evaluation}
\label{sec:weak}\index{Weak coupling}
Suppose we regard $\lambda$ as a small parameter, so let us expand
(\ref{energy}) in powers of $\lambda$.  The first term is
\be
\mathcal{E}^{(1)}=-\frac\lambda{8\pi a^2}\sum_{m=-\infty}^\infty \int_0^\infty
\D x\,x^2\,\frac{\D}{\D x}K_m(x)I_m(x).\label{1storder}
\ee
The addition theorem for the modified Bessel functions is
\be
K_0(kP)=\sum_{m=-\infty}^\infty \E^{\I m(\phi-\phi')}K_m(k\rho)I_m(k\rho'),\quad
\rho>\rho',\label{addthm}
\ee
where $P=\sqrt{\rho^2+\rho^{\prime2}-2\rho\rho'\cos(\phi-\phi')}$. If this
is extrapolated to the limit $\rho'=\rho$ we conclude that the sum of the
Bessel functions appearing in (\ref{1storder}) is $K_0(0)$, that is, a
constant, so there is no first-order contribution to the energy.
For a rigorous derivation of this result, see \cite{CaveroPelaez:2006rt}.

We can proceed the same way to evaluate the second-order contribution,
\be
\mathcal{E}^{(2)}=\frac{\lambda^2}{16\pi a^2}\int_0^\infty \D x\,x^2\,
\frac{\D}{\D x}\sum_{m=-\infty}^\infty I_m^2(x)K_m^2(x).\label{2ndorden}
\ee
By squaring the sum rule (\ref{addthm}), and taking the limit
$\rho'\to\rho$, we evaluate the sum over Bessel functions appearing here
as
\be
\sum_{m=-\infty}^\infty I_m^2(x)K_m^2(x)=\int_0^{2\pi}\frac{\D\varphi}{2\pi}
K_0^2(2x\sin\varphi/2).\label{squaresr}
\ee
Then changing the order of integration we find that the second-order energy can be
written as
\be
\mathcal{E}^{(2)}=-\frac{\lambda^2}{64\pi^2 a^2}\int_0^{2\pi}\frac{\D\varphi}
{\sin^2\varphi/2}\int_0^\infty \D z\,z\,K_0^2(z),\label{phive2}
\ee
where the Bessel-function integral has the value 1/2.  However, the integral
over $\varphi$ is divergent.  We interpret this integral by adopting
 an analytic regularization based on the integral \cite{Cavero-Pelaez:2004xp}
\be
\int_0^{2\pi}\D\varphi \left(\sin\frac\varphi2\right)^s=\frac{2\sqrt{\pi}\Gamma
\left(\frac{1+s}2\right)}{\Gamma\left(1+\frac{s}2\right)},\label{sinint}
\ee
which holds for $\mbox{Re}\,s>-1$.  Taking the right-side of this equation
to define the $\varphi$ integral for all $s$, we conclude that the
$\varphi$ integral in (\ref{phive2}),
and hence the second-order energy $\mathcal{E}^{(2)}$, is zero.

\subsubsection{Numerical Evaluation}
\label{sec:num}
Given that the above argument evidently formally omits divergent terms, it may
be more satisfactory, as in \cite{Cavero-Pelaez:2004xp}, to offer a
numerical evaluation of $\mathcal{E}^{(2)}$. (The corresponding argument
for $\mathcal{E}^{(1)}$ is given in \cite{CaveroPelaez:2006rt}.) We can very efficiently do so
using the uniform asymptotic expansions (\ref{uae}).
Thus the asymptotic behavior of the product of Bessel functions appearing
in (\ref{2ndorden}) is
\be
I_m^2(x)K_m^2(x)\sim\frac{t^2}{4m^2}\left(1+\sum_{k=1}^\infty \frac{r_k(t)}
{m^{2k}}\right).\label{i2k2}
\ee
The first three polynomials occurring here are
\alpheqn\label{r}
\bea
r_1(t)&=&\frac{t^2}4(1-6t^2+5t^4),\label{r1}\\
r_2(t)&=&\frac{t^4}{16}(7-148t^2+554 t^4-708 t^6+295 t^8),\label{r2}\\
r_3(t)&=&\frac{t^6}{16}(36-1666t^2+13775t^4-44272t^6\nonumber\\
&&\quad\mbox{}+67162t^8-48510t^{10}
+13475 t^{12}).\label{r3}
\eea
\reseteqn

We now write the second-order energy (\ref{2ndorden}) as
\bea
\mathcal{E}^{(2)}&=&-\frac{\lambda^2}{8\pi a^2}\Bigg\{\int_0^\infty
\D x\,x\left[I_0^2(x)K_0^2(x)-\frac1{4(1+x^2)}\right]\nonumber\\
&&\quad\mbox{}-\frac14\lim_{s\to 0} \left( \frac 1 2 +
\sum_{m=1}^\infty m^{-s}\right)\int_0^\infty \D z
\,z^{2-s}\frac{\D}{\D z}\frac1{1+z^2}\nonumber\\
&&\quad\mbox{}+
2\int_0^\infty \D z\,z\frac{t^2}4\sum_{m=1}^\infty \sum_{k=1}^3 \frac{r_k(t)}
{m^{2k}}\nonumber\\
&&\quad\mbox{}+2\sum_{m=1}^\infty
\int_0^\infty \D x\,x\left[I_m^2(x)K_m^2(x)-\frac{t^2}{4m^2}
\left(1+\sum_{k=1}^3\frac{r_k(t)}{m^{2k}}\right)\right]\Bigg\}.\label{num}
\eea
In the final integral $z=x/m$. The successive terms are evaluated as
\bea
\mathcal{E}^{(2)}&\approx&-\frac{\lambda^2}{8\pi a^2}\Bigg[
\frac14(\gamma+\ln4)-\frac14\ln2\pi-\frac{\zeta(2)}{48}
+\frac{7\zeta(4)}{1920}-\frac{31\zeta(6)}{16128}\nonumber\\
&&\quad+0.000864+0.000006\Bigg]=-\frac{\lambda^2}{8\pi a^2}(0.000000),
\eea
where  in the last term in (\ref{num}) only the $m=1$ and 2 terms are
significant. Therefore, we have demonstrated numerically
that the energy in order $\lambda^2$ is zero to an accuracy of better than
$10^{-6}$.

The astute reader will note that we used a standard, but possibly
questionable, analytic regularization in defining the second term in
(\ref{num}), where the initial sum and integral are only
defined for $1<s<2$, and then the result is continued to $s=0$.
Alternatively, we could follow \cite{Cavero-Pelaez:2004xp} and insert
there an exponential regulator in each integral of $\E^{-x\delta}$, with
$\delta$ to be taken to zero at the end of the calculation.  For $m\ne0$
$x$ becomes $mz$, and then the sum on $m$ becomes
\be
\sum_{m=1}^\infty \E^{-mz\delta}=\frac1{\E^{z\delta}-1}.\label{sumonm}
\ee
Then when we carry out the integral over $z$ we obtain for that term
\be
\frac\pi{8\delta}-\frac14\ln2\pi.\label{cutoff1}
\ee
Thus we obtain the same finite part as above, but in addition an explicitly
divergent term
\be
\mathcal{E}^{(2)}_{\rm div}=-\frac{\lambda^2}{64 a^2\delta}.
\ee
If we think of the cutoff in terms of a vanishing proper time $\tau$,
$\delta=\tau/a$, this divergent term is proportional to $1/a$, so the
divergence in the energy goes like $L/a$, if $L$ is the (very large) length
of the cylinder.  This is of the form of the shape divergence encountered
in \cite{Cavero-Pelaez:2004xp}.

\subsubsection{Divergences in the Total Energy}
\label{hk}
In this subsection we are going to use heat-kernel knowledge to
determine the divergence structure in the total energy. We
consider a general cylinder of the type ${\cal C} = \mathbb{R} \times
Y$, where $Y$ is an arbitrary smooth two dimensional region rather
than merely being the disc. As a metric we have $\D s^2 = \D z^2 +
\D Y^2$ from which we obtain that the zeta function (density)
associated with the Laplacian on ${\cal C}$ is ($\mbox{Re}\,s>3/2$)
\bea
 \zeta (s) &=& \frac 1 {2\pi} \int_{-\infty}^\infty \D k
 \sum_{\lambda_Y} (k^2 + \lambda_Y)^{-s} = \frac 1 {2\pi}
 \frac{\sqrt \pi \Gamma \left( s-\frac 1 2 \right)}{\Gamma (s)}
 \sum _{\lambda_Y} \lambda_Y ^{1/2 -s} \nn\\
&=& \frac 1 {2\pi}
 \frac{\sqrt \pi \Gamma \left( s-\frac 1 2 \right)}{\Gamma (s)} \zeta_Y
\left(s-\frac 1 2 \right) .\eea
Here $\lambda_Y$ are the eigenvalues of the Laplacian on $Y$, and
$\zeta_Y (s)$ is the zeta function associated with these
eigenvalues. In the zeta-function scheme the Casimir energy is
defined as \index{Zeta-function regularization}
\be \left. E_{\rm Cas} = \frac 1 2 \mu^{2s} \,\,\zeta
\left( s-\frac 1 2 \right) \right|_{s=0}, \ee
which, in the
present setting, turns into
\bea \left. E_{\rm Cas} = \frac 1 2
\mu^{2s} \frac {\Gamma (s-1)} {2\sqrt \pi \Gamma \left( s- \frac 1
2 \right)} \zeta_Y (s-1) \right|_{s=0}. \eea
Expanding this
expression about $s=0$, one obtains
\be E_{\rm Cas} = \frac 1 {8\pi
s} \zeta_Y (-1) + \frac 1 {8\pi} \left( \zeta _Y (-1) \left[ 2 \ln
(2\mu ) -1\right] + \zeta_Y ' (-1) \right) + {\cal O } (s).\ee
The contribution associated with $\zeta _Y (-1)$ can be
determined solely from the heat-kernel coefficient knowledge,
namely
\be \zeta _Y (-1) = - a_4, \ee
in terms of the standard 4th heat-kernel coefficient.\index{Heat kernel 
coefficients} The contribution coming
from $\zeta _Y ' (-1)$ can in general not be determined.
But as we see, at least the divergent term can be determined entirely by
the heat-kernel coefficient.

The situation considered
in the Casimir energy calculation is a $\delta$-function shell
along some smooth line $\Sigma$ in the plane (here, a
circle of radius $a$). The manifolds considered
are the cylinder created by the region inside of the line, and
the region outside of the line; from the results the
contribution from free Minkowski space has to be subtracted to avoid
trivial volume divergences (the representation in terms of the
Bessel functions already has Minkowski space contributions
subtracted). The $\delta$-function shell generates a jump in the
normal derivative of the eigenfunctions;
call the jump $U$ (here, $U=\lambda/a$).
The leading heat-kernel coefficients for
this situation, namely for functions which are continuous
across the boundary but which have
a jump of the first normal derivative at the boundary, have been
determined in \cite{gilkey}; the
relevant $a_4$ coefficient is given in Theorem 7.1, p.~139 of that
reference. The
results there are very general; for our purpose there is exactly one
term that survives, namely
\be a_4 = - \frac 1 {24 \pi}
\int_\Sigma \D l\, U^3, \ee
 which shows that
\be E_{\rm Cas}^{\rm div} =
\frac 1 {192 \pi^2 s}\int_\Sigma \D l\, U^3. \ee
So no matter
along which line the $\delta$-function shell is concentrated, the
first two orders in a weak-coupling expansion do not contribute
any divergences in the total energy.
But the third order does, and the divergence is given above.

For the example considered, as mentioned, $U=\lambda / a$ is
constant, and the integration leads to the length of the line
which is $2\pi a$. Thus we get for this particular
example
\be \mathcal{E}_{\rm Cas}^{\rm div} = \frac 1 {96 \pi s}
\frac{\lambda^3}{a^2}. \label{lambda3div}
\ee
[Compare this with the corresponding divergence for a sphere,
(\ref{secondhk}).]
This can be easily checked from the
explicit representation we have for the energy. We have already
seen that the first two orders in $\lambda$ identically vanish, while
the part of the third order that potentially contributes a
divergent piece is
\be\mathcal{E}^{(3)}= - \frac 1 {8 \pi a^2} \sum_{m=-\infty}
^\infty \int_0^\infty \D x\, x^{2-2s} \frac \D {\D
x} \frac 1 3\lambda^3 K_m ^3 (x) I_m ^3 (x).\label{pdp} \ee
The $m=0$ contribution is well behaved about $s=0$; while for the remaining sum
using
\be K_m^3 (mz) I_m ^3 (mz) \sim \frac 1 {8m^3} \frac 1
{(1+z^2)^{3/2} },\ee
 we see that the leading contribution is
\bea
\mathcal{E}^{(3)}&\sim&-\frac{\lambda^3}{12 \pi a^2} \sum_{m=1}^\infty m^{2-2s}
\int_0^\infty \D z \,z^{2-2s} \frac \D {\D z} \frac
1 {8m^3} \frac 1 {(1+z^2)^{3/2}} \nn\\
&=& - \frac {\lambda^3}{96 \pi
a^2} \zeta_R (1+2s) \int_0^\infty \D z\, z^{2-2s} \frac
\D {\D z} \frac 1 {(1+z^2)^{3/2}} \nn\\
& =&\frac {\lambda^3}{96 \pi a^2} \zeta_R
(1+2s)\frac{\Gamma (2-s) \Gamma \left( s+\frac 1 2 \right)}
{\Gamma (3/2)} = \frac {\lambda^3}{96 \pi a^2 s} + {\cal O} (s^0),
\eea in perfect agreement with the heat-kernel prediction (\ref{lambda3div}).

\subsection{Strong Coupling}
\label{sec:sc} \index{Casimir energy for Dirichlet cylinder}
The strong-coupling limit of the energy (\ref{energy}), that is, the
Casimir energy of a Dirichlet cylinder,
\be
\mathcal{E}^D=-\frac1{8\pi a^2}\sum_{m=-\infty}^\infty \int_0^\infty \D x\,x^2
\frac{\D}{\D x}\ln I_m(x)K_m(x),
\ee
was worked out to high accuracy by Gosdzinsky and Romeo \cite{gosrom},
\be
\mathcal{E}^D=\frac{0.000614794033}{a^2}.\label{grresult}
\ee
It was later redone with less accuracy by Nesterenko and Pirozhenko
\cite{nest}.

For completeness, let us sketch the evaluation here.
We carry out a numerical calculation (very similar to that of \cite{nest})
in the spirit of Sect.~\ref{sec:num}.  We add and subtract the leading
uniform asymptotic expansion (for $m=0$ the asymptotic behavior) as follows:
\bea
\mathcal{E}^D&=&-\frac1{8\pi a^2}\Bigg\{-2\int_0^\infty \D x\,x\left[
\ln\left(2xI_0(x)K_0(x)\right)-\frac18\frac1{1+x^2}\right]\nonumber\\
&&\quad\mbox{}+2\sum_{m=1}^\infty \int_0^\infty \D x\,x^2\frac{\D}{\D x}
\left[\ln \left(2x I_m(x)K_m(x)
\right)-\ln\left(\frac{xt}{m}\right)-\frac12\frac{r_1(t)}{m^2}\right]
\nonumber\\
&&\quad\mbox{}-2\left(\frac12+\sum_{m=1}^\infty\right)
\int_0^\infty \D x\,x^2\frac{\D}{\D x}\ln 2x
+2\sum_{m=1}^\infty \int_0^\infty \D x\,x^2 \frac{\D}{\D x}\ln xt
\nonumber\\
&&\quad\mbox{}+\sum_{m=1}^\infty \int_0^\infty \D x\,x^2 \frac{\D}{\D x}\left[
\frac{r_1(t)}{m^2}-\frac14\frac1{1+x^2}\right]\nonumber\\
&&\quad\mbox{}+\frac14\left(\frac12+\sum_{m=1}^\infty\right)
\int_0^\infty \D x\,x^2\frac{\D}{\D x}\frac{1}{1+x^2}\Bigg\}.
\label{scints}\eea
In the first two terms we have subtracted the leading asymptotic behavior so
the resulting integrals are convergent.  Those terms are restored in the
fourth, fifth, and sixth terms.  The most divergent part of the Bessel
functions
are removed by the insertion of $2x$ in the corresponding integral, and
its removal in the third term.  (As we've seen
above, such terms have been referred to
as ``contact terms,'' because if a time-splitting
regulator, $\E^{\I \zeta\tau}$, is inserted into the
frequency integral, a term proportional to $\delta(\tau)$ appears,
which is zero as long as $\tau\ne0$.)  The terms involving Bessel functions are evaluated
numerically, where it is observed that the asymptotic value of the
summand (for large $m$)
in the second term is $1/32m^2$.  The fourth term is evaluated by writing it
as
\be
2\lim_{s\to0}\sum_{m=1}^\infty m^{2-s}\int_0^\infty \D z
\frac{z^{1-s}}{1+z^2}=2\zeta'(-2)=-\frac{\zeta(3)}{2\pi^2},\label{4th}
\ee
while the same argument, as anticipated, shows that the third ``contact'' term
is zero,\footnote{This argument is a bit suspect, since the
analytic continuation that defines the integrals has no common region
of existence.  Thus the argument in the following subsection may be
preferable.  However, since that term is properly a contact term,
it should in any event be spurious.} while the sixth term is
\be
-\frac12\lim_{s\to 0}\left[\zeta(s)+\frac12\right]\frac1s=\frac14\ln 2\pi.
\ee
The fifth term is elementary.  The result then is
\bea
\mathcal{E}^D&=&\frac1{4\pi a^2}\left(0.010963-0.0227032+0+0.0304485+0.21875
-0.229735\right)\nonumber\\
&=&\frac1{4\pi a^2}(0.007724)=\frac{0.0006146}{a^2},\label{myresult}
\eea
which agrees with (\ref{grresult}) to the fourth significant figure.

\subsubsection{Exponential Regulator}
As in Sect.~\ref{sec:num}, it may seem more satisfactory to insert an
exponential regulator rather than use analytic regularization.  Now it is
the third, fourth, and sixth terms in (\ref{scints}) that must be treated.
The latter is just the negative of (\ref{cutoff1}).  We can combine the
third and fourth terms to give using (\ref{sumonm})
\be
-\frac1{\delta^2}-\frac2{\delta^2}\int_0^\infty \frac{\D z\,z^3}{z^2+\delta^2}
\frac{\D^2}{\D z^2}\frac{1}{\E^z-1}.
\ee
The latter integral may be evaluated by writing it as an integral along
the entire $z$ axis, and closing the contour in the upper half plane,
thereby encircling the poles at $i\delta$ and at $2in\pi$, where $n$ is a
positive integer.  The residue theorem then gives for that integral
\be
-\frac{2\pi}{\delta^3}-\frac{\zeta(3)}{2\pi^2},
\ee
so once again we obtain the same finite part as in (\ref{4th}).
In this way of proceeding, then, in addition to the finite part in
(\ref{myresult}), we obtain divergent terms
\be
\mathcal{E}^D_{\rm div}=\frac1{64a^2\delta}+\frac1{8\pi a^2\delta^2}
+\frac1{4a^2\delta^3},
\ee
which, with the previous interpretation for $\delta$, implies divergent
terms in the energy proportional to $L/a$ (shape), $L$ (length), and
$aL$ (area), respectively.  Such terms presumably are to be subsumed
in a renormalization of parameters in the model.  Had a logarithmic
divergence occurred [as does occur in weak coupling in
$\mathcal{O}(\lambda^3)$] such a
renormalization would apparently be impossible---however, see
\cite{CaveroPelaez:2006rt}. \index{Renormalization}

\subsection{Local Energy Density}
\label{sec:local}
We compute the energy density from the stress tensor (\ref{st2}), or
\bea
\langle T^{00}\rangle&=&\frac1{2\I}\left(\partial^0\partial^{0\prime}
+\bnabla\cdot\bnabla'\right)G(x,x')
\bigg|_{x'=x}-\frac\xi{\I}\nabla^2G(x,x)
\nonumber\\
&=&\frac1{16\pi^3\I}\int_{-\infty}^\infty \D k\int_{-\infty}^\infty \D\omega
\sum_{m=-\infty}^\infty \Bigg[\left(\omega^2+k^2+\frac{m^2}{r^2}+\partial_r
\partial_{r'}\right)g(r,r')\bigg|_{r'=r}\nonumber\\
&&\quad\mbox{}-2\xi\frac1r\partial_r r\partial_r g(r,r)\Bigg].
\eea
We omit the free part of the Green's function, since that corresponds to
the energy that would be present in the vacuum in the absence of
the cylinder. When we
insert the remainder of the Green's function (\ref{gout2}), we obtain
the following expression for the energy density outside the cylindrical
shell:
\bea
u(r)&=&\langle T^{00}-T_{(0)}^{00}\rangle=-\frac\lambda{16\pi^3}
\int_{-\infty}^\infty
\D\zeta\int_{-\infty}^\infty \D k\sum_{m=-\infty}^\infty \frac{
I_m^2(\kappa a)}{1+\lambda I_m(\kappa a)K_m(\kappa a)}
\nonumber\\
&&\times\left[\left(2\omega^2+\kappa^2+\frac{m^2}{r^2}\right)K_m^2(\kappa
r)+\kappa^2K_m^{\prime 2}(\kappa r)-2\xi\frac1r\frac\partial{\partial r}
r\frac\partial{\partial r} K_m^2(\kappa r)\right],\nonumber\\
&&\qquad\qquad r>a.\label{uofr}
\eea
The factor in square brackets can be easily seen to be, from the
modified Bessel equation,
\be
2\omega^2 K_m^2(\kappa r)+\frac{1-4\xi}2\frac1r\frac\partial{\partial r}
r\frac\partial{\partial r} K_m^2(\kappa r).\label{9.3}
\ee
For the interior region, $r<a$, we have the corresponding expression for
the energy density with $I_m\leftrightarrow K_m$.

\subsection{Total and Surface Energy}
We first need to verify that we recover the expression for the energy
found in Sect.~\ref{sec:pressure}.  So let us integrate expression (\ref{uofr})
over the region exterior of the cylinder, and the corresponding
interior expression over the inside region. The second term in (\ref{9.3})
is a total derivative, while the first is exactly the one evaluated in
Sec.~\ref{sec:pressure}.  The result is
\bea
2\pi\int_0^\infty \D r\,r\,u(r)&=&-\frac1{8\pi a^2}\sum_{m=-\infty}^\infty
\int_0^\infty \D x\,x^2\frac{\D}{\D x}
\ln\left[1+\lambda I_m(x)K_m(x)\right]\nonumber\\
&&\mbox{}-(1-4\xi)\frac\lambda{4\pi a^2}
\int_0^\infty \D x\,x
\sum_{m=-\infty}^\infty \frac{I_m(x)K_m(x)}{1+\lambda I_m(x)K_m(x)}.\nonumber\\
\label{inten}
\eea
The first term is the total energy (\ref{energy}), but what do we make of
the second term?  In strong coupling, it would represent a constant that
should have no physical significance (a contact term---it is independent
of $a$ if we revert to the physical variable $\kappa$ as the integration
variable).  In general, however,
there is another contribution to the total energy, residing precisely
on the singular surface.  This surface energy is given in general
by \cite{dowkerandkennedy,Kennedy:1979ar,saharian,Romeo:2001dd,
Fulling:2003zx,Milton:2004vy} \index{Surface energy}
\be
\hat E=-\frac{1-4\xi}{2i}\oint_S \D\mathbf{S}\cdot\bnabla G(x,x')
\bigg|_{x'=x},\label{surfen}
\ee
as given for $\xi=0$ in (\ref{es1}),
where the normal to the surface is out of the region in question.
In this case it is easy to see that $\hat E$ exactly equals the
negative of the second term in (\ref{inten}).  This is an example of the
general theorem (\ref{einclst})
\be
\int(\D\mathbf{r}) u(\mathbf{r})+\hat E=E,\label{surfthm}
\ee
that is, the total energy $E$
is the sum of the integrated local energy density
and the surface energy.  The generalization of this theorem,
(\ref{surfen}) and (\ref{surfthm}), to curved space is given in
\cite{Saharian:2003dr}.
A consequence of this theorem is that the
total energy, unlike the local energy density, is independent of
the conformal parameter $\xi$.
(Note that this surface energy vanishes when $\xi=1/4$ as Fulling
has stressed \cite{Fulling:2008ha}.)

\subsection{Surface Divergences}
We now turn to an examination of the behavior of the local energy
density (\ref{uofr}) as $r$ approaches $a$ from outside the cylinder.
To do this we use the uniform asymptotic expansion (\ref{uae}).  
Let us begin by
considering the strong-coupling limit, a Dirichlet cylinder.  If we stop with
only the leading asymptotic behavior, we obtain the expression
\bea
u(r)&\sim&-\frac1{8\pi^3}\int_0^\infty \D\kappa\,\kappa\,2\sum_{m=1}^\infty
\E^{-m\chi}\Bigg\{\left[-\kappa^2+(1-4\xi)\left(\kappa^2+\frac{m^2}{r^2}\right)
\right]\frac{\pi t}{2m}\nonumber\\
&&\qquad\mbox{}+(1-4\xi)\kappa^2\frac{\pi}{2mt}\frac1{z^2}
\Bigg\},\qquad (\lambda\to\infty),
\eea
where
\be
\chi=-2\left[\eta(z)-\eta\left(z\frac{a}r\right)\right],
\ee
and we have replaced the integral over $k$ and $\zeta$ by one over the
polar variable $\kappa$ as before.
Here we ignore the difference between $r$ and $a$ except in the exponent, and
we now replace $\kappa$ by $m z/a$.  Close to the surface,
\be
\chi\sim \frac2t\frac{r-a}r,\quad r-a\ll r,
\ee
and we carry out the sum over $m$ according to
\be
2\sum_{m=1}^\infty m^3 e^{-m\chi}\sim-2\frac{\D^3}{\D\chi^3}\frac1\chi
=\frac{12}{\chi^4}\sim\frac34\frac{t^4r^4}{(r-a)^4}.
\ee
Then the energy density behaves, as $r\to a+$,
\bea
u(r)&\sim&-\frac3{64\pi^2}\frac1{(r-a)^4}\int_0^\infty \D z\,z[t^5+t^3(1-8\xi)]
\nonumber\\
&=&-\frac1{16\pi^2}\frac1{(r-a)^4}(1-6\xi).
\eea
This is the universal surface divergence first discovered by Deutsch
and Candelas \cite{deutsch} and seen for the sphere in (\ref{dsphere1}) 
\cite{Cavero-Pelaez:15kq}.  It therefore occurs, with precisely the
same numerical coefficient, near a Dirichlet plate \cite{Milton:2002vm}.  
Unless gravity is considered, it is utterly without physical
significance, and may be eliminated with the conformal choice for the
parameter $\xi$, $\xi=1/6$.

We will henceforth make this conformal choice.  Then the leading divergence
depends upon the curvature.  This was also worked out by Deutsch and
Candelas \cite{deutsch}; for the case of a cylinder, that result is
\be
u(r)\sim \frac1{720\pi^2}\frac1{r(r-a)^3},\quad r\to a+,\label{dccyl}
\ee
exactly 1/2 that for a Dirichlet sphere of radius $a$ (\ref{sdsc})
\cite{Cavero-Pelaez:15kq}, as anticipated from the general analysis
summarized in (\ref{divstructure}).
Here, this result may be straightforwardly derived by keeping the
$1/m$ corrections in the uniform asymptotic expansion (\ref{uae}),
as well as the next term in the expansion of $\chi$,
(\ref{chiexp}).

\subsubsection{Weak Coupling}
\index{Weak coupling}
Let us now expand the energy density (\ref{uofr}) for small coupling,
\bea
u(r)&=&-\frac{\lambda}{16\pi^3}\int_{-\infty}^\infty \D\zeta
\int_{-\infty}^\infty \D k\sum_{m=-\infty}^\infty I_m^2(\kappa a)
\sum_{n=0}^\infty(-\lambda)^n I_m^n(\kappa a)K_m^n(\kappa a)\nonumber\\
&&\quad\times\left\{\left[-\kappa^2+(1-4\xi)\left(\kappa^2+\frac{m^2}{r^2}
\right)
\right]K_m^2(\kappa r)+(1-4\xi)\kappa^2 K_m^{\prime2}(\kappa r)\right\}.\nn\\
\eea
If we again use the leading uniform asymptotic expansions for the Bessel
functions, we obtain the expression for the leading behavior of the term
of order $\lambda^{n}$,
\be
u^{(n)}(r)\sim \frac1{8\pi^2r^4}\left(-\frac\lambda2\right)^{n}
\int_0^\infty \D z\,z\sum_{m=1}^\infty m^{3-n}\E^{-m\chi}t^{n-1}(t^2+1-8\xi).
\ee
The sum on $m$ is asymptotic to
\be
\sum_{m=1}^\infty m^{3-n}\E^{-m\chi}\sim (3-n)!\left(\frac{t r}{2(r-a)}
\right)^{4-n},\quad r\to a+,\label{summ}
\ee
so the most singular behavior of the order $\lambda^n$ term is, as $r\to a+$,
\be
u^{(n)}(r)\sim (-\lambda)^n\frac{(3-n)!\,(1-6\xi)}{96\pi^2 r^n(r-a)^{4-n}}.
\ee
This is exactly the result found for the weak-coupling limit for a
$\delta$-sphere (\ref{surfdivn}) \cite{Cavero-Pelaez:15kq} and for 
a $\delta$-plane (\ref{pertsd}) \cite{Milton:2004vy},
so this is also a universal result, without physical significance.  It may be
made to vanish by choosing the conformal value $\xi=1/6$.

With this conformal choice, once again we must expand to higher order.
We use the corrections noted above, in (\ref{uae}) and
(\ref{chiexp}), (\ref{tztd}).
Then again a quite simple calculation gives
\be
u^{(n)}\sim(-\lambda)^n\frac{(n-1)(n+2)\Gamma(3-n)}{2880\pi^2 r^{n+1}
(r-a)^{3-n}},\quad r\to a+,
\ee
which is analytically continued from the region $1\le \mbox{Re}\,
n<3$.  Remarkably, this
is exactly one-half the result found in the same weak-coupling expansion
for the leading conformal divergence outside a sphere (\ref{sdwc})
\cite{Cavero-Pelaez:15kq}.
Therefore, like the strong-coupling result (\ref{dccyl}),
this limit is universal, depending on the sum of the principal curvatures
of the interface.

In \cite{CaveroPelaez:2006rt} we considered a annular shell of finite
thickness, which as the thickness $\delta$ tended to zero gave
a finite residual energy in the annulus, in terms of the energy
density $u$ in the annulus,
\be
\mathcal{E}_{\rm ann}=2\pi\delta a u\sim(1-4\xi)\frac{\lambda}{4\pi a^2}
\sum_{m=-\infty}^\infty \int_0^\infty \D\kappa a\,\kappa a
\frac{I_m(\kappa a)K_m(\kappa a)}{1+\lambda I_m(\kappa a)K_m(\kappa a)}=
\hat{\mathcal{E}},\label{shellissurf}
\ee
which is exactly the form of the surface energy given by the negative
of the second term in (\ref{inten}).
In particular, note that the term in $\hat{\mathcal{E}}$ of order $\lambda^3$
is, for the conformal value $\xi=1/6$, exactly equal to that term in the
total energy $\mathcal{E}$ (\ref{energy}): [see (\ref{pdp})]
\be
\hat{\mathcal{E}}^{(3)}=\mathcal{E}^{(3)}.
\ee
This means that the divergence encountered in the global
energy (\ref{lambda3div})
is exactly accounted for by the divergence in the surface energy, which
would seem to provide strong evidence in favor of the renormalizablity
of that divergence. \index{Renormalization}\index{Surface energy}

\section{Gravitational acceleration of Casimir energy}

We will here show that a body undergoing uniform acceleration
(hyperbolic motion) imparts the same acceleration to the
quantum vacuum energy associated with this body.  This
is consistent with the equivalence principle that states
that all forms of energy should gravitate equally.  A
general variational argument, which, however, did not deal
with the divergent parts of the energy, was given in
\cite{Fulling:2007xa}. This section is based on \cite{Milton:2007ar}.

\subsection{Green's Functions in Rindler Coordinates}
\label{cg}\index{Acceleration of Casimir energy}
\index{Rindler coordinates}
\index{Gravitational acceleration}
Relativistically, uniform acceleration is described by hyperbolic
motion,
\be
z = \xi \cosh \tau\quad\mathrm{and} \quad t = \xi \sinh \tau.
\label{hy-eq}
\ee
Here the proper acceleration of the particle described by these
equations is $\xi^{-1}$, and we have
chosen coordinates so that at time $t=0$, $z(0)=\xi$.
Here we are going to consider the corresponding metric
\begin{equation}
\D s^2 = - \D t^2 + \D z^2 + \D x^2 + \D y^2
     = - \xi^2 \D\tau^2 + \D\xi^2 + \D x^2 + \D y^2.
\end{equation}
In these coordinates, the d'Alembertian operator takes on cylindrical form
\be
-\left(\frac\partial{\partial t}\right)^2
+\left(\frac\partial{\partial z}\right)^2+\bnabla_\perp^2
=-\frac1{\xi^2}\left(\frac\partial{\partial\tau}\right)^2+\frac1\xi
\frac\partial{\partial\xi}\left(\xi\frac\partial{\partial\xi}\right)
+\bnabla_\perp^2,
\ee
where $\perp$ refers to the $x$-$y$ plane.

\subsubsection{Green's Function for One Plate}
For a scalar field in these coordinates, subject to a potential $V(x)$,
the action is
\begin{equation}
W = \int \D^4x \sqrt{-g(x)}
{\cal{L}}(\phi(x)),\label{action}
\end{equation}
where $x\equiv(\tau,x,y,\xi)$ represents the coordinates,
$\D^4x = \D\tau \,\D\xi\, \D x\, \D y$ is the coordinate volume element,
$g_{\mu\nu}(x) = \mbox{diag}(-\xi^2,+1,+1,+1)$ defines the metric,
$g(x) = \det g_{\mu\nu}(x)=-\xi^2$ is the determinant of the metric,
and the Lagrangian density is
\begin{equation}
{\cal{L}}(\phi(x))
= - \frac{1}{2} g_{\mu\nu}(x) \partial^\mu \phi(x)\partial^\nu \phi(x)
  - \frac{1}{2} V(x) \phi(x)^2,
\end{equation}
where for a single semitransparent plate located at $\xi_1$
\begin{equation}
V(x) = \lambda \delta(\xi - \xi_1),
\end{equation}
and $\lambda>0$ is the coupling constant having dimensions of mass.
More explicitly we have
\begin{equation}
W = \int \D^4x \,\frac{\xi}{2}
\left[\frac{1}{\xi^2}
 \left( \frac{\partial \phi}{\partial \tau} \right)^2
- \left( \frac{\partial \phi}{\partial \xi} \right)^2
- \left( {\bnabla}_\perp \phi \right)^2
- V(x) \phi^2
\right].
\end{equation}
Stationarity of the action under an arbitrary variation in
the field leads to the equation of motion
\begin{equation}
\left[
- \frac{1}{\xi^2} \frac{\partial^2}{\partial \tau^2}
+ \frac{1}{\xi} \frac{\partial}{\partial \xi}
  \xi \frac{\partial}{\partial \xi}
+ {\bnabla}_\perp^2 - V(x)
\right] \phi(x) = 0.\label{eom}
\end{equation}

The corresponding Green's function satisfies the differential equation
\begin{equation}
-\left[
- \frac{1}{\xi^2} \frac{\partial^2}{\partial \tau^2}
+ \frac{1}{\xi} \frac{\partial}{\partial \xi}
  \xi \frac{\partial}{\partial \xi}
+ {\bnabla}_\perp^2 - V(x)
\right] G(x,x^\prime) =
\frac{\delta(\xi-\xi^\prime)}{\xi}
\delta(\tau - \tau^\prime) \delta({\bf x}_\perp-{\bf x}_\perp^\prime).
\end{equation}
Since in our case $V(x)$ has only $\xi$ dependence we
can write this in terms of the reduced Green's function $g(\xi, \xi')$,
\begin{equation}
G(x,x^\prime)
= \int_{-\infty}^{\infty}
  \frac{\D\omega}{2 \pi} \int \frac{\D^2 \mathbf{k}_\perp}{(2 \pi)^2}
  \E^{-\I \omega (\tau - \tau^\prime)}
  \E^{\I {\bf k}_\perp \cdot ({\bf x} - {\bf x}^\prime)_\perp}
  g(\xi,\xi^\prime),
\label{capG}
\end{equation}
where $g(\xi,\xi^\prime)$ satisfies
\begin{equation}
-\left[
\frac{1}{\xi} \frac{\partial}{\partial \xi}
  \xi \frac{\partial}{\partial \xi}
+\frac{\omega^2}{\xi^2}
- k_\perp^2 - V(x)
\right] g(\xi,\xi^\prime) = \frac{\delta(\xi-\xi^\prime)}{\xi}.
\label{zgf}
\end{equation}

We recognize this equation as defining the semitransparent cylinder
problem discussed in Sect.~\ref{sec:gf}
\cite{CaveroPelaez:2006rt}, with the replacements
\be
a\to\xi_1,\quad
m\to\zeta=-i\omega,\quad \kappa\to k=k_\perp,\quad \lambda\to\lambda\xi_1,
\ee
so that from (\ref{gin2}) and (\ref{gout2})
we may immediately write down the solution in terms of modified
Bessel functions,
\alpheqn
\bea
g(\xi,\xi')&=&I_\zeta(k\xi_<)K_\zeta(k\xi_>)
-\frac{\lambda\xi_1 K_\zeta^2(k\xi_1)
I_\zeta(k\xi)I_\zeta(k\xi')}{1+\lambda\xi_1 I_\zeta(k\xi_1)K_\zeta(k\xi_1)},
\qquad \xi,\xi'<\xi_1,\nn\\ \label{gfrind1}\\
&=&I_\zeta(k\xi_<)K_\zeta(k\xi_>)-\frac{\lambda \xi_1 I_\zeta^2(k\xi_1)
K_\zeta(k\xi)K_\zeta(k\xi')}{1+\lambda\xi_1 I_\zeta(k\xi_1)K_\zeta(k\xi_1)},
\qquad \xi,\xi'>\xi_1.\nn\\ \label{gfrind2}
\eea
\reseteqn
Note that in the strong-coupling limit, $\lambda\to\infty$, this reduces
to the Green's function satisfying Dirichlet boundary conditions at
$\xi=\xi_1$.

\subsubsection{Minkowski-space Limit}

To recover the Minkowski-space Green's function for the semitransparent plate,
we use the uniform asymptotic expansion (Debye expansion), based on the
limit
\bea
\xi\to\infty,&&\quad \xi_1\to\infty, \quad \xi-\xi_1 \mbox{ finite },
\quad \zeta\to\infty,\quad \zeta/\xi_1 \mbox{ finite }.
\eea
For large $\zeta$ we use (\ref{uae}) with $x=\zeta z=k\xi$, for example.
Expanding the above expressions (\ref{gfrind1}), (\ref{gfrind2}) 
around some arbitrary point $\xi_0$, chosen such that the differences
$\xi - \xi_0$, $\xi'-\xi_0$, and $\xi_1 - \xi_0$ are finite, we
find for the leading term, for example,
\begin{equation}
\sqrt{\xi \xi'} \, I_\zeta(k\xi) K_\zeta(k\xi')
\sim \frac{1}{2 \kappa} \, \E^{\kappa (\xi - \xi')},
\label{IKap}
\end{equation}
where $\kappa^2 = k^2 + \hat\zeta^2$,  $\hat\zeta=\zeta/\xi_0$.
In this way, taking for simplicity $\xi_0=\xi_1$, we find the Green's function
for a single plate in Minkowski space,
\begin{equation}
\xi_1g(\xi,\xi')\to g^{(0)}(\xi,\xi^\prime) =
\frac{1}{2 \kappa} \,\E^{-\kappa |\xi - \xi^\prime|}
- \frac{\lambda}{\lambda + 2 \kappa}
  \frac{1}{2 \kappa}
  \, \E^{-\kappa|\xi - \xi_1|} \E^{-\kappa|\xi^\prime - \xi_1|}.
\label{1-g0-zzp}
\end{equation}

\subsubsection{Green's Function for Two Parallel Plates}\label{sec2c}
For two semitransparent plates perpendicular to the $\xi$-axis
and located at $\xi_1$, $\xi_2$, with couplings $\lambda _1$ and
$\lambda_2$, respectively, we find the following form for
the Green's function:
\alpheqn
\bea
g(\xi,\xi')&=&I_<K_>
-\frac{\lambda_1\xi_1K_1^2+\lambda_2\xi_2K_2^2-\lambda_1\lambda_2\xi_1\xi_2
K_1K_2(K_2I_1-K_1I_2)}{\Delta}II_\prime,\nn\\ &&\qquad\qquad\xi, \xi'<\xi_1, \\
&=&I_<K_>-\frac{\lambda_1\xi_1I_1^2
+\lambda_2\xi_2I_2^2+\lambda_1\lambda_2\xi_1\xi_2
I_1I_2(I_2K_1-I_1K_2)}{\Delta}KK_\prime,\nn\\ &&\qquad\qquad\xi,\xi'>\xi_2, \\
&=&I_<K_>
-\frac{\lambda_2\xi_2K_2^2(1+\lambda_1\xi_1K_1I_1)}\Delta II_\prime\nn\\
&&\quad\mbox{}-\frac{\lambda_1\xi_1I_1^2(1+\lambda_2\xi_2K_2I_2)}
\Delta KK_\prime
+\frac{\lambda_1\lambda_2\xi_1\xi_2I_1^2K_2^2}\Delta(IK_\prime+KI_\prime),\nn\\
&&\qquad\qquad \xi_1<\xi,\xi'<\xi_2,
\eea
\reseteqn
where
\be
\Delta
=(1+\lambda_1\xi_1K_1I_1)(1+\lambda_2\xi_2K_2I_2)-\lambda_1\lambda_2\xi_1\xi_2
I_1^2K_2^2,\label{Delta}
\ee
and we have used the abbreviations
$I_1=I_\zeta(k\xi_1)$, $I=I_\zeta(k\xi)$, $I_\prime=I_\zeta(k\xi')$, etc.

Again we can check that these formulas reduce to the well-known
Minkowski-space limits.  In the $\xi_0\to\infty$ limit,
the uniform asymptotic expansion (\ref{uae}) gives, for $\xi_1<\xi,\xi'<\xi_2$
\bea
\xi_0g(\xi,\xi')&\to& g^{(0)}(\xi,\xi')=\frac1{2\kappa}\E^{-\kappa|\xi-\xi'|}
+\frac1{2\kappa\tilde\Delta}\bigg[\frac{\lambda_1\lambda_2}
{4\kappa^2}2\cosh\kappa(\xi-\xi')\nn\\
&&\mbox{}-\frac{\lambda_1}{2\kappa}\left(1+\frac{\lambda_2}{2\kappa}\right)
\E^{-\kappa(\xi+\xi'-2\xi_2)}
-\frac{\lambda_2}{2\kappa}\left(1+\frac{\lambda_1}{2\kappa}\right)
\E^{\kappa(\xi+\xi'-2\xi_1)}\bigg],\nn\\
\eea
where ($a=\xi_2-\xi_1$)
\be
\tilde\Delta=\left(1+\frac{\lambda_1}{2\kappa}\right)
\left(1+\frac{\lambda_2}{2\kappa}\right)\E^{2\kappa a}
-\frac{\lambda_1\lambda_2}{4\kappa^2},\label{tDelta}
\ee
which is exactly the expected result (\ref{gin}), (\ref{delta0}).
The correct limit is also obtained in the other two regions.

\subsection{Gravitational Acceleration of Casimir Apparatus}

We next consider the situation when the plates are forced to
``move rigidly'' \cite{born} in such a way that the proper distance
between the plates is preserved. This is achieved if the
two plates move with different but
constant proper accelerations.

The canonical energy-momentum or stress tensor derived from the
action (\ref{action}) is \index{Stress tensor}
\begin{equation}
T_{\alpha \beta}(x)
= \partial_\alpha \phi(x) \partial_\beta \phi(x)
+ g_{\alpha \beta}(x) {\cal{L}}(\phi(x)),
\end{equation}
where the Lagrange density includes the $\delta$-function potential. The
components referring to the pressure and the energy density are
\alpheqn
\begin{eqnarray}
T_{33}(x)
&=&  \frac{1}{2}\frac{1}{\xi^2}
 \left( \frac{\partial \phi}{\partial \tau} \right)^2
+ \frac{1}{2} \left( \frac{\partial \phi}{\partial \xi} \right)^2
- \frac{1}{2} \left( {\bnabla}_\perp \phi \right)^2
- \frac{1}{2} V(x) \phi^2,
\\
\frac{1}{\xi^2} \, T_{00}(x)
&=&  \frac{1}{2}\frac{1}{\xi^2}
 \left( \frac{\partial \phi}{\partial \tau} \right)^2
+ \frac{1}{2} \left( \frac{\partial \phi}{\partial \xi} \right)^2
+ \frac{1}{2} \left( {\bnabla}_\perp \phi \right)^2
+ \frac{1}{2} V(x) \phi^2.
\end{eqnarray}
\reseteqn
The latter may be written in an alternative convenient form using
the equations of motion (\ref{eom}):
\be
T_{00}=\frac12\left(\frac{\partial\phi}{\partial\tau}\right)^2-\frac12\phi
\frac{\partial^2}{\partial\tau^2}\phi+\frac\xi 2\frac\partial{\partial\xi}
\left(\phi\xi\frac\partial{\partial\xi}\phi\right)
+\frac{\xi^2}2\bnabla_\perp\cdot(\phi\bnabla_\perp\phi),
\label{enden}
\ee
which is the appropriate version of (\ref{t00pluseom}) here.
The force density is given by
\be
f_\lambda=-\frac1{\sqrt{-g}}\partial_\nu(\sqrt{-g}T^\nu{}_\lambda)
+\frac12T^{\mu\nu}\partial_\lambda g_{\mu\nu},
\ee
or in Rindler coordinates
\be
f_\xi=-\frac1\xi\partial_\xi(\xi T^{\xi\xi})-\xi T^{00}.\label{fxi}
\ee
When we integrate over all space to get the force, the first term is
a surface term which does not contribute:\footnote{Note that in previous
works, such as \cite{Milton:2004vy,Milton:2004ya}, the surface
term was included, because the integration was carried out only over
the interior and exterior regions.  Here we integrate over the surface
as well, so the additional so-called surface energy is automatically
included.  This is described in the argument leading to (\ref{es1}).
  Note, however, if (\ref{fxi}) is integrated over a
small interval enclosing the $\delta$-function potential,
$$
\int_{\xi_1-\epsilon}^{\xi_1+\epsilon} \D\xi\,\xi f_\xi=-\xi_1\Delta T^{\xi\xi},
$$
where $\Delta T^{\xi\xi}$ is the discontinuity in the normal-normal
component of the stress density.  Dividing this expression by $\xi_1$ gives
the usual expression for the force on the plate.}
\be
\mathcal{F}=\int \D\xi\, \xi f_\xi=-\int\frac{\D\xi}{\xi^2}T_{00}.
\label{ep}
\ee
This could be termed the Rindler coordinate force per area, defined as the
change in momentum per unit Rindler coordinate time $\tau$ per unit
cross-sectional area.  If we
multiply $\mathcal{F}$  by the gravitational acceleration $g$ we obtain
the gravitational force per area on the Casimir energy.
This result (\ref{ep}) seems entirely consistent with the equivalence
principle, since $\xi^{-2}T_{00}$ is the energy density.
Using the expression (\ref{enden}) for the energy density,
taking the vacuum expectation value, and rescaling
$\zeta=\hat\zeta\xi$,
we see that the gravitational force per cross sectional area is merely
\be
\mathcal{F}=\int \D\xi \,\xi\int\frac{\D\hat\zeta \, \D^2\mathbf{k}}
{(2\pi)^3}\hat\zeta^2g(\xi,\xi).\label{gravf-gf}
\ee

This result for the energy contained in the force equation (\ref{gravf-gf})
is an immediate consequence of the general formula
for the Casimir energy (\ref{casenergy}) \cite{miltonbook}.

Alternatively, we can start from the following formula for the force
density for a single semitransparent plate, following directly from
the equations of motion (\ref{eom}),
\be
f_\xi=\frac12\phi^2\partial_\xi \lambda\delta(\xi-\xi_1).\label{fd}
\ee
The vacuum expectation value of this yields the force
in terms of the Green's function,
\be
\mathcal{F}=-\lambda\frac12
\int\frac{\D\zeta\,\D^2\mathbf{k}}
{(2\pi)^3}\partial_{\xi}[\xi g(\xi,\xi)]\bigg|_{\xi=\xi_1}.
\ee

\subsubsection{Gravitational Force on a Single Plate}

For example, the force on a single plate at $\xi_1$ is given by
\be
\mathcal{F}=-\partial_{\xi_1} \frac12\int\frac{\D\zeta\,\D^2\mathbf{k}}
{(2\pi)^3}\ln[1+\lambda \xi_1 I_\zeta(k\xi_1)K_\zeta(k\xi_1)],
\ee
Expanding this about some arbitrary point $\xi_0$, with $\zeta=\hat\zeta\xi_0$,
using the uniform asymptotic expansion (\ref{uae}), we get ($\kappa^2
=k^2+\hat\zeta^2$)
\be
\xi_1 I_\zeta(k\xi_1)K_\zeta(k\xi_1)\sim\frac{\xi_1}{2\zeta}\frac1
{\sqrt{1+(k\xi_1/\zeta)^2}}\approx\frac{\xi_1}{2\kappa\xi_0}\left(1-\frac{k^2}
{\kappa^2}\frac{\xi_1-\xi_0}{\xi_0}\right).
\ee
From this, if we introduce polar coordinates for the $\mathbf{k}$-$\hat\zeta$
integration, the
coordinate force is
\bea
\mathcal{F}&=&-\frac12\partial_{\xi_1}\frac{\xi_0}{2\pi^2}\int_0^\infty
\D\kappa\,\kappa^2\frac{\lambda}{2\kappa+\lambda}
\left(1+\frac{\xi_1-\xi_0}{\xi_0}
\right)\left(1-\frac{\langle k^2\rangle}{\kappa^2}\frac{\xi_1-\xi_0}{\xi_0}
\right)\nn\\
&=&-\frac\lambda{4\pi^2}\partial_{\xi_1}(\xi_1-\xi_0)\int_0^\infty
\frac{\D\kappa}{2\kappa+\lambda}\langle\hat\zeta^2\rangle
\nn\\
&=&-\frac{1}{96\pi^2a^3}\int_0^\infty\frac{\D y\,y^2}{1+y/\lambda a},
\label{singleplate}
\eea
where for example
\be
\langle\hat\zeta^2\rangle=\frac12\int_{-1}^1 \D\cos\theta\, \cos^2\theta\,
\kappa^2=\frac13\kappa^2.
\ee
The divergent expression (\ref{singleplate})
 is just the negative of the quantum vacuum energy of a single plate,
seen in (\ref{31energy}) and (\ref{eslab}).


\subsubsection{Parallel Plates Falling in a Constant Gravitational Field}

In general, we have two alternative forms for the
gravitational force on the two-plate system:
\be
\mathcal{F}=-(\partial_{\xi_1}+\partial_{\xi_2})\frac12\int\frac{\D\zeta\,
\D^2\mathbf{k}}{(2\pi)^3}\ln\Delta,\ee
$\Delta$ given in (\ref{Delta}),
which is equivalent to (\ref{gravf-gf}).  (In the latter, however,
bulk energy, present if no plates are present, must be omitted.)
From either of the above two methods, we find the coordinate force
is given by
\be
\mathcal{F}=-\frac1{4\pi^2}\int_0^\infty \D\kappa\,\kappa^2 \ln\Delta_0,
\label{coordf}
\ee
where $\Delta_0=\E^{-2\kappa a}\tilde\Delta$, $\tilde\Delta$ given in
(\ref{tDelta}).
The integral may be easily shown to be
\alpheqn
\bea
\mathcal{F}&=&\frac1{96\pi^2 a^3}\int_0^\infty \D y\,y^3\frac{1+
\frac1{y+\lambda_1a}
+\frac1{y+\lambda_2a}}{\left(\frac{y}{\lambda_1a}+1\right)
\left(\frac{y}{\lambda_2a}+1\right)e^y-1}\nn\\
&&\quad\mbox{}-\frac1{96\pi^2 a^3}\int_0^\infty \D y\,y^2\left[\frac1{\frac{y}{\lambda_1a}
+1}+\frac1{\frac{y}{\lambda_2a}+1}\right]\label{negce}\\
&=&-(\mathcal{E}_c+\mathcal{E}_{d1}+\mathcal{E}_{d2}),\label{fisen}\eea
\reseteqn
which is just the negative of the Casimir energy of the two semitransparent
plates including the divergent pieces---See (\ref{31energy})
 \cite{Milton:2004vy,Milton:2004ya}.
Note that $\mathcal{E}_{di}$, $i=1,2$, are simply the divergent energies
(\ref{singleplate}) associated with a single plate.

\subsubsection{Renormalization}
\index{Renormalization}
The divergent terms in (\ref{fisen})
simply renormalize the masses (per unit area) of each plate:
\bea
E_{\rm total}&=&m_1+m_2+\mathcal{E}_{d1}+\mathcal{E}_{d2}+\mathcal{E}_c\nn\\
&=&M_1+M_2+\mathcal{E}_c,
\eea
where $m_i$ is the bare mass of each plate, and the renormalized mass
is $M_i=m_i+\mathcal{E}_{di}$.
Thus the gravitational force on the entire apparatus obeys the
equivalence principle
\be
g\mathcal{F}=-g(M_1+M_2+\mathcal{E}_c).
\ee
The minus sign reflects the downward acceleration of gravity on the
surface of the earth.  Note here that the Casimir interaction energy
$\mathcal{E}_c$ is negative, so it reduces the gravitational attraction
of the system.

\subsection{Summary}

We have found, in conformation with the result given in
\cite{Fulling:2007xa}, an extremely simple answer to the question of
how Casimir energy accelerates in a weak gravitational field:
Just like any other form of energy, the gravitational force $F$ divided
by the area of the plates is
\be
\frac{F}A=-g\mathcal{E}_c.
\ee
This is the result expected by the equivalence principle, but
is in contradiction to some earlier disparate claims in the
literature \cite{karim,calloni,caldwell,Sorge:2005ed,bimonte}.
Bimonte et al.~\cite{Bimonte:2008zva} now agree completely with
our conclusions.
This result perfectly agrees with that found by Saharian
et al.~\cite{Saharian:2003fd}
for Dirichlet, Neumann, and perfectly conducting plates for the finite
Casimir interaction energy.  The acceleration of Dirichlet plates follows from
our result when the strong coupling limit $\lambda\to\infty$ is taken.
What makes our conclusion particularly interesting is that it refers not
only to the finite part of the Casimir interaction energy between
semitransparent plates, but to the divergent parts as well, which are
seen to simply renormalize the gravitational mass of each plate, as they
would the inertial mass.  The reader may object that by equating gravitational
force with uniform acceleration we have built in the equivalence principle,
and so does any procedure based on Einstein's equations; but the real
nontriviality here is that quantum fluctuations obey the same universal
law.  The reader is also referred to the important work on this subject
by Jaekel and Reynaud \cite{Jaekel:2008bs}, and extensive references
therein.


\section{Conclusions}

In this review, I have illustrated the issues involved in calculating
self-energies in the simple context of massless scalar fields interacting with
$\delta$-function potentials, so-called semitransparent boundaries.
This is not as unrealistic as it might sound, since in the strong coupling
limit this yields Dirichlet boundary conditions, and by using derivative
of $\delta$-function boundaries, we can recover Neumann boundary conditions.
Thus, where the boundaries admit the separation into TE and TM modes, we
can recover perfect-conductor boundaries imposed on electromagnetic fields.

We have examined both divergences occurring in the total energy, and 
divergences which appear in the local energy density as boundaries are 
approached.  The latter divergences often have little to do with the former,
because the local divergences may cancel across the boundaries, and they
typically depend on the form (canonical or conformal, for example) of
the local stress-energy tensor.  The global divergences apparently can
always be uniquely isolated, leaving a unique finite self-energy; in some
cases at least the divergent parts can be absorbed into a renormalization
of properties of the boundaries, such as their masses.  It is expected that
if the ideal boundaries were represented as a solitonic structure arising
from a background field, this ``renormalization'' idea could be put on a
more rigorous footing. \index{Renormalization} 

Evidence for the consistency of this view occurs in the parallel plate
configuration, where we show that the finite interaction energy and the
divergent self-energies of each plate exhibit the same inertial and 
gravitational properties, that is, are each consistent with the equivalence
principle.  Thus it is indeed consistent to absorb the self-energies into
the masses of each plate.  We hope to prove in the future that this 
renormalization consistency is a general feature.

In spite of the length of this review, we have barely scratched the surface.
In particular, we have not discussed how the divergent contributions of
the local stress tensor are consistent with Einstein's equations
\cite{Estrada:2008zz}.  We have also only discussed simple separable 
geometries, where the equations for the Green's functions can be solved
on both the inside and the outside of the boundaries.  This excludes
the extensive work on rectangular cavities, where only the sum over
interior eigenvalues can be carried out \cite{lukosz,lukosz1,lukosz2,
ambjorn,Actor:1995mc}.  There are some numerical coincidences, for
example between the energy for a sphere and a cube, but since divergences
have been simply omitted by zeta-function regularization, the significance
of the latter results remains unclear.  There are a few other examples
where the interior Casimir contribution can be computed exactly, while
the exterior problem cannot be solved, an example being a cylinder with
cross section of an equilateral triangle.  Such results seem more problematic
than those we have discussed here.  

We also have not discussed semiclassical and numerical techniques.
For example, there is the extremely interesting work of Schaden
\cite{Schaden:2006qg}, who computes a very accurate approximation
for the Casimir energy of a spherical shell using optical path techniques.
The same technique gives zero for the cylindrical shell, not the attractive
value found in \cite{DeRaad:1981hb}, which is not surprising.
Not unrelated to this technique is the exact worldline method of
Gies and collaborators \cite{Gies:2006xe,Gies:2006cq,Gies:2006bt}, which
is able to capture edge effects. The optical path work of
Scardicchio and Jaffe \cite{Jaffe:2003mb,Scardicchio:2004fy,Schroeder:2004iv}
should also be cited, although it is largely restricted to examining
the forces between distinct bodies.
 This review also does not refer to
the remarkable progress in numerical techniques, some of which are
related to the multiple scattering approach---for some recent
references see \cite{Graham:2009zb,Rahi:2007qt}, and the contributions to
this volume by Emig, Jaffe, and Rahi and by Johnson---, which however,
have not yet been turned to examining self-interactions.

The central issue is the meaning of Casimir self-energy, and how, in
principle, it might be observed.  Probably the right direction to address
such issues is in terms of quantum corrections to solitons---for example, see
\cite{Farhi:1998vx,Graham:1998qq,CaveroPelaez:2009vi}.  The issues
being considered go to the very heart of renormalized quantum
field theory, and likely to the meaning and origin of mass, 
a subject about which we in fact know very little.

\begin{acknowledgement}
I thank the US Department of Energy and the US National Science Foundation
for partial support of this work.  I thank my many collaborators, 
including Carl Bender, Iver Brevik, In\'es Cavero-Pel\'aez,
 Lester DeRaad, Steve Fulling, Ron Kantowski, Klaus Kirsten,
Vladimir Nesterenko, Prachi Parashar, August Romeo,  K.V. Shajesh,
and Jef Wagner, for
their contributions to the work described here.
 
\end{acknowledgement}
%

%
%
%

\end{document}